\documentclass{elsart3p}

\usepackage{graphicx}

\usepackage{amssymb}

\usepackage{epsfig}
\usepackage{amsmath}
\usepackage{xspace}

\usepackage{ifpdf}
\usepackage{calc}
\usepackage{comment}
\usepackage{url}

\newcommand{\NuSTAR}{{\it NuSTAR}\xspace}
\newcommand{\pe}[1]{{\it ProtoEXIST{#1}}\xspace}

\newcommand{\sS}[1]{\mbox{$\rm{}^{#1}$}}
\newcommand{\Ss}[1]{\mbox{$\rm{}_{#1}$}}

\newcommand{\Swift}{\mbox{$Swift$}\xspace}
\newcommand{\EXIST}{\mbox{$EXIST$}\xspace}
\newcommand{\Deg}{\mbox{$^\circ$}\xspace}
\newcommand{\x}{\mbox{$\times$}\xspace}
\newcommand{\arcmin}{\mbox{$^\prime$}\xspace}

\begin{document}

\begin{frontmatter}



\title{Flight Performance of an advanced CZT Imaging Detector in \\
 	a Balloon-borne Wide-Field Hard X-ray Telescope -- \it ProtoEXIST1
	\thanksref{cor}}
\thanks[cor]{Send correspondence to J. Hong
	: \\
	E-mail: jaesub@head.cfa.harvard.edu \\
}


\author{J.~Hong, B.~Allen, J.~Grindlay}
\address{Harvard-Smithsonian Center for Astrophysics, Cambridge, MA 02138}
\author{S.~Barthelemy, R.~Baker}
\address{NASA Goddard Space Flight Center, Greenbelt, MD 20771}
\author{A.~Garson, H.~Krawczynski}
\address{Washington University in St. Louis and the McDonnell Center
for the Space Sciences, St. Louis, MO 63130}
\author{J.~Apple, W.~H.~Cleveland}
\address{NASA Marshall Space Flight Center and Universities Space Research Association, Huntsville, AL 35812}

\begin{abstract}

We successfully carried out the first high-altitude balloon flight of a
wide-field hard X-ray coded-aperture telescope \pe1, which was launched
from the Columbia Scientific Balloon Facility at Ft. Sumner, New Mexico
on October 9, 2009.  \pe1 is the first implementation of an advanced
CdZnTe (CZT) imaging detector in our ongoing program to establish the technology
required for next generation wide-field hard X-ray telescopes such as the
High Energy Telescope (HET) in the Energetic X-ray Imaging Survey Telescope
(\EXIST).  The CZT detector plane in \pe1 consists of an 8 \x 8 array of
closely tiled 2 cm \x 2 cm \x 0.5 cm thick pixellated CZT crystals, each with 8 \x 8 pixels, 
mounted on a set of readout electronics boards and covering a 256
cm\sS{2} active area with 2.5 mm pixels.  A tungsten mask, mounted at
90 cm above the
detector provides shadowgrams of X-ray sources in the 30 -- 600 keV band
for imaging, allowing a fully coded field of view of 9\Deg \x 9\Deg (and
19\Deg \x 19\Deg for 50\% coding fraction) with an angular resolution of 20\arcmin.
In order to reduce the background radiation, the detector is surrounded
by semi-graded (Pb/Sn/Cu) passive shields on the four
sides  all the way to the mask. On the back side, a 26 cm \x 26 cm
\x 2 cm CsI(Na) active shield provides signals to tag
charged particle induced events as well as $\gtrsim$ 100 keV background
photons from below. The flight duration was only about
7.5 hours due to strong winds (60 knots) at float altitude (38--39 km).
Throughout the flight, the CZT detector performed excellently. The telescope
observed Cyg X-1, a bright black hole binary system,
for $\sim$ 1 hour at the end of the flight. 
Despite a few problems with the pointing and aspect systems that caused the
telescope to track about 6.4 deg
off the target, the analysis of the Cyg X-1 data
revealed an X-ray source at 7.2$\sigma$ in the 30--100 keV
energy band at the expected location  from the optical images
taken by the onboard daytime star camera.  The success of this first
flight is very encouraging for the future development of the advanced
CZT imaging detectors (\pe2, with 0.6 mm pixels), which will take advantage of the
modularization architecture employed in \pe1.

\end{abstract}

\end{frontmatter}

\section{Introduction}

The high energy sky is active and dynamic, full of inherently variable
sources as well as extremely luminous Gamma-ray Bursts
(GRBs). Wide-field hard X-ray imaging is perhaps the only effective way to
capture these unpredictable, energetic celestial events. The current GRB
mission, \Swift, to study GRBS, employs a large area (0.5 m\sS{2})
array of CdZnTe (CZT) detectors, each 4 mm \x 4 mm \x 2 mm as separate
pixels for coded-aperture imaging.
The previously proposed \EXIST telescope would have had 4.5 m\sS{2} with
0.6 mm pixels \cite{Grindlay10}. A smaller (1024 cm\sS{2}) CZT imager 
incorporating the same 0.6mm pixels has now been proposed 
for the MIRAX experiment on the {\it Lattes} mission to be launched by Brazil
\cite{Grindlay11}.
The combination of a large area and fine pixels imposes
serious challenges in detector design, packaging, and operations, given the 
limited power and space available in space missions.

We have been pushing the technology of advanced CZT imaging detectors
required for future generation of wide-field hard X-ray imagers through
our ongoing, three-part, balloon-borne wide-field hard X-ray telescope
experiment, \pe.  We have successfully carried out a first high-altitude
balloon flight of \pe1, the first implementation of the program, which
was launched from the NASA Columbia Scientific Balloon Facility (CSBF)
at Ft. Sumner, New Mexico on October 9, 2009.  \pe1 is a wide-field
hard X-ray coded-aperture telescope, built to develop an efficient
modularization and packaging architecture for a
large area fine pixel CZT imaging detector (256 cm\sS{2} with 2.5 mm
pixels) and to demonstrate its performance in a near space environment. The
X-ray telescope including the CZT detector performed very well during
the flight.  Here we report the detailed performance of the X-ray
telescope and the imaging CZT detector during the flight.
Full details of the detector and
telescope system, including the assembly of the coded mask, the shields,
the onboard calibration source and the interface to gondola will be
presented in separate papers
\cite{Allen10, Allen11}. 

\section{\pe1 Instrument}

\begin{figure} \begin{center}
\includegraphics*[width=0.37\textwidth]{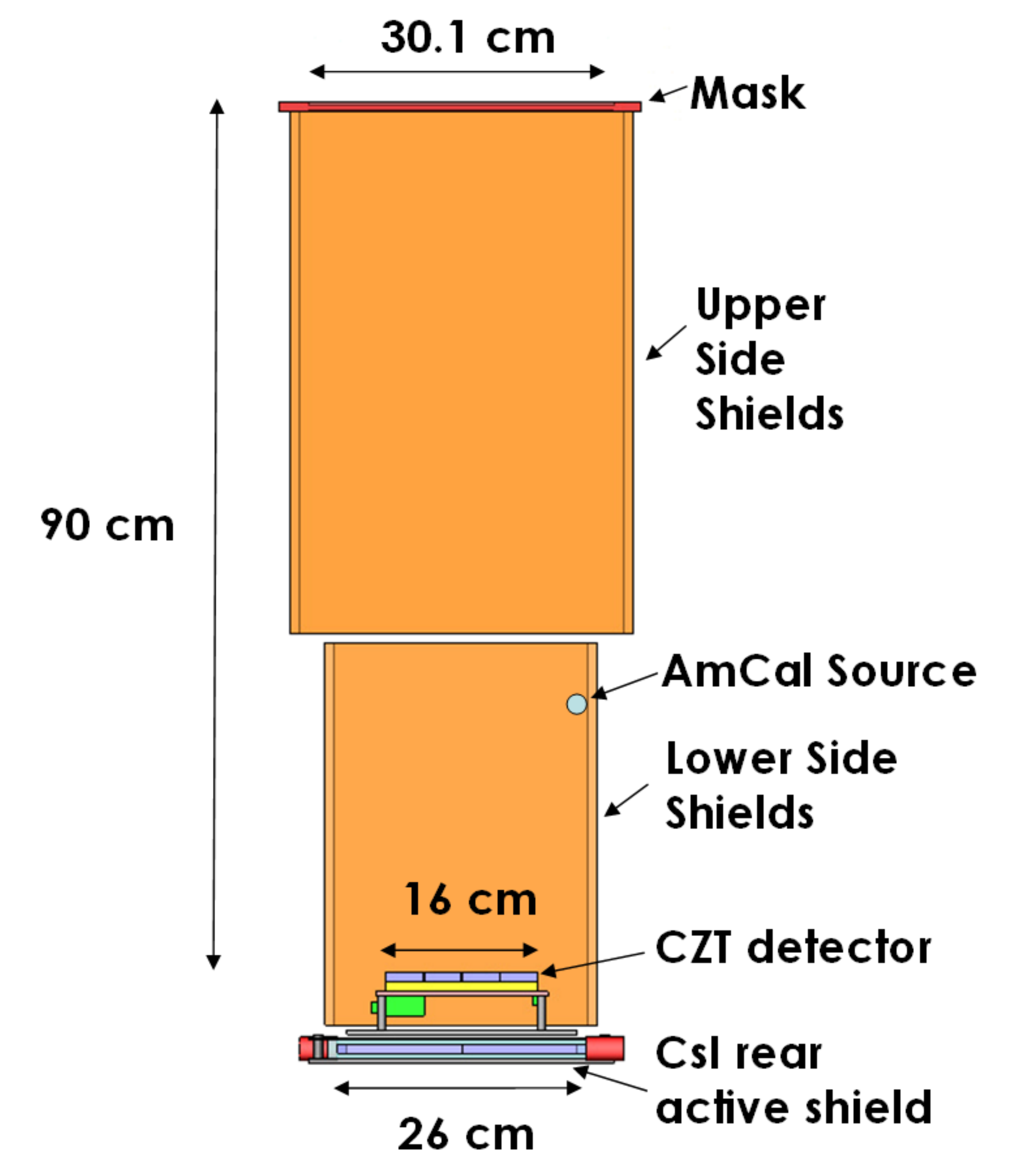}
\caption{Schematic view of \pe1. The mask and the upper side shields
are mounted on the top cover of the pressure vessel (PV), and the rest
of the system is mounted inside the PV. See also Table~\ref{telescope}.}
\label{pe1}
\end{center}
\end{figure}

\pe1 consists of a CZT detector plane, a Tungsten mask, passive side
shields and a CsI rear active shield. 
The schematic view of the \pe1 telescope is
shown in Fig~\ref{pe1}.  The key parameters of \pe1 are
summarized in Table 1.

\subsection{CZT detector plane}

The CZT detector in \pe1 is a first implementation of a close-tiled
fine pixel imaging CZT array in our program.
Fig.~\ref{detector} shows the detector assembly and the fully integrated detector
plane in the flight configuration.
An initial laboratory performance
and the assembly procedure of the
CZT detector plane for \pe1 are reported in \cite{Hong09} in detail. 
Here we review the modularization and packaging architecture and
operational modes briefly.

The full detector plane covers 16 \x 16 cm\sS{2} with 8 \x 8 CZT crystals,
providing 64 \x 64 imaging pixels with 2.5 mm pitch. All of the 64 CZT
crystals are tiled with a uniform gap ($\sim 900$ $\mu$m for 19.5 mm
\x 19.5 mm CZT), allowing the total active area of 15.6 \x 15.6 cm\sS{2}. 
This is a factor of $\gtrsim$ 8 times larger in area than any
previous tiled-pixellated CZT imagers of which we are aware.

The basic building block is a Detector Crystal Unit (DCU) that
consists of a 1.95 cm \x 1.95 cm \x 0.5 cm CZT crystal with 8 \x
8 pixels bonded to an interposer board with a RadNET ASIC \cite{Cook98}
underneathfor readout (Fig.~\ref{detector}a).  The traces on an
interposer board
connect the 2-D array of 8 \x 8 anode pixel pads on a CZT to a 1-D array
of 64 channel input pads of the RadNET ASIC.  In the next stage of assembly, 
a 2 \x 4 array of DCUs are mounted on two vertically stacked electronics
boards to make a Detector Crystal Array (DCA) (Fig.~\ref{detector}b). The
DCA electronics boards contain four Analog-to-Digital Convertors
(ADCs, AD7685BRM) and a Field Programmable Gate Array (FPGA, Altera Cyclone II
EP2C20F256C7). The FPGA is programmed to have four independent data
processing channels, each of which handles two DCUs with one shared ADC.
We closely tile 2 \x 4 DCAs on a motherboard, the FPGA-Controller Board
(FCB), to complete the detector plane or Detector
Module (DM; Fig.~\ref{detector}c).  The FCB contains another larger FPGA
(Altera Cyclone II EP2C20F484C7) to process 32 independent data streams
(1 per 2 DCUs) from the FPGAs of the 8 DCAs.
A network card
(NetBurner\footnote{\url{http://www.netburner.com/}}, MOD5282)
is mounted on the FCB to collect the science data from the FPGA and other
HouseKeeping (HK) information, and pass them to a flight computer through
an Ethernet port (see \S\ref{s:fc} and {\cite{Allen11}).
The FCB also reserves 8 TTL input lines for event tagging with
\sS{241}Am Calibration (AmCal, 200 nCi) and active shield trigger signals 
and for absolute timing using 1 pulse per second (PPS) pulses from
the GPS unit on the gondola.

\begin{table}
\small
\caption{Telescope Parameters of \pe1}
\begin{tabular*}{0.48\textwidth}{l@{\extracolsep{\fill}}l}
\hline\hline
Parameters			&	Values 		\\
\hline
Sensitivity (5 $\sigma$) 	&	$\sim$ 140 mCrab/hr\sS{a}\\
Energy Range			&	30 -- 300 keV\sS{b} 	\\
Energy Resolution		&	3 -- 5 keV	\\
Field of View 			&	19\Deg \x 19\Deg (50\% Coding)\\
				&	9\Deg \x 9\Deg (Fully Coded)\\
Angular Resolution		&	20.3$^\prime$ 	\\
\hline
CZT Detector			&	64 crystals \x (2 \x 2 \x 0.5 cm\sS{3}) \\
Area 				&	16 \x 16 cm\sS{2} 	\\
Pixel, Thickness		&	2.5 mm, 5 mm 	\\
\hline
Tungsten Mask 			&	12 layers \x 0.3 mm thick \\
Coding Area			&	30.1 \x 30.1 cm\sS{2}	\\
Pixel, Grid, Thickness		&	4.7 mm, 0.5 mm, 3.6 mm 	\\
\hline
Mask-Det. Separation		&	90 cm 	\\
CsI(Na) Rear Shield		&	26 \x 26 \x 2 cm\sS{3}	\\
Side Passive Shield		&	Semi-graded Pb/Sn/Cu	\\
\hline
\sS{241}Am Cal. Source		&	200 nCi, 36 cm above the detector	\\
\hline
\end{tabular*}
(a) assuming a 86\% QE and a 45\% dead time fraction with an observed
$\sim$ 250 cps background rate in the 30--200 keV band (see \S\ref{s:flight} and
Fig.~\ref{ctimeline}b), on-axis and no atmospheric absorption.
(b) for imaging with $>$80\% mask efficiency as limited by mask
thickness; the detector energy range is $\sim$ 30 -- 600 keV.
\label{telescope}
\end{table}

We obtain CZT crystals from Redlen
Technologies\footnote{\url{http://www.redlen.com}}, which produce high
quality CZT at low cost through their patented traveling heater method.
We bias 64 CZTs at $-$600V using two
EMCO\footnote{\url{http://www.emchohighvoltage.com/}} High Voltage Power
Supplies (HV PS, C12N) mounted on the FCB. We apply the HV lead on the cathode
using a thin (50 $\mu$m) Al tape, which is suitable for pressurized balloon
flights and convenient for replacing DCU units if needed. The Redlen CZTs show
$<$ 0.5 nA/pixel leakage currents under $-$600V. Each DCU is surrounded
by Kapton-coated Copper RFI shields to minimize the observed interference
between the units (see \cite{Hong09}). The shields kept the
electronics noise down to the level of a single DCU operation.

\begin{figure*} \begin{center}
\includegraphics*[width=0.90\textwidth]{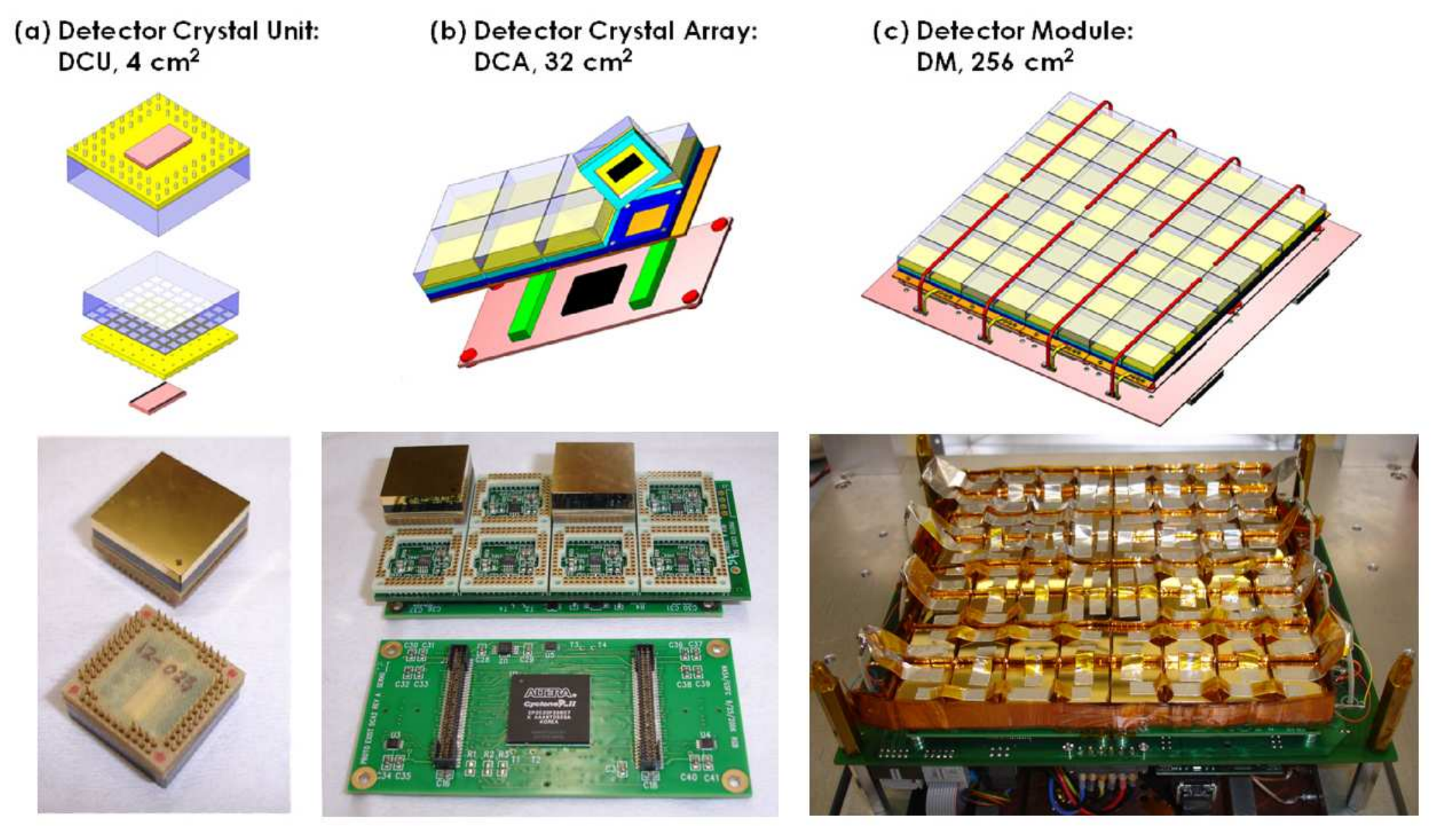}
\caption{The CZT detector assembly for the \pe1 imaging detector.
(a) Detector Crystal Unit (DCU) consists of 2 \x 2 cm\sS{2} \x 0.5 cm CZT
bonded via an interposer board to the RadNET ASIC with 1-D 64 inputs
(pink). (b)
Detector Crystal Array (DCA) consists of a 2x4 array of DCUs. (c)
Final view of flight Detector Module (DM) covers 16 \x 16 cm\sS{2} with a 2
\x 4 array of DCAs. HV bias ($-$600V) lines connect to the 4 DCAs on
each side through Al tape "jumpers" between DCUs.
}
\label{detector}
\end{center}
\end{figure*}

CZTs were bonded on an interposer board (IPB) with two low-cost,
low-temperature ($\lesssim$ 100 {\Deg}C) techniques -- initially with
low temperature solder
bonds from Delphon/QuikPak\footnote{\url{http://quikpak.com}} and
then with Transient Liquid Phase Sintering (TLPS) bonds from Creative
Electron Inc\footnote{\url{http://creativeelectron.com}} for final
DCUs. TLPS bonding was explored for possible future applications and
one of its main advantages is that the resulting bonding is re-workable:
the CZT can be detached and re-attached if needed.

The RadNet ASIC was chosen for \pe1 mainly because of its relatively low
power consumption (100 -- 150 $\mu$W/pixel) and the expected advance of
the subsequent ASIC members in the same development family
\cite{Cook98,Harrison10}.  The low power consumption is an essential
requirement for a large area CZT detector with fine pixels, which
results in a large number of pixels or channels (e.g. 256k pixels
in MIRAX).  In addition, the RadNet ASIC provides a wide dynamic
range with low electronics noise suitable for the hard X-ray band and
allows flexible readout modes -- multi-pixel pulse profile readout -- for
depth sensing (based on neighbor pixel charges, \cite{Hong06,Luke94}) 
and event reconstruction
of multi-pixel triggers arising
from split charges or Compton scattering \cite{Hong09}.

For \pe1, we have implemented four readout modes for debugging, trigger,
normal operation, or calibration (see Table 1 in \cite{Hong09}).
During the flight, we operate the detector in the normal mode, where
we read signals from eight neighbor pixels and two reference pixels in
addition to the trigger pixels.  The total number of neighbor pixels
to read out varies, depending on the location of the trigger pixels:
five for edge pixels and three for corner pixels. In \pe1, we
have not recorded the signals from adjacent
crystals since the electrostatic shields prevent the charge induction
across the crystals.

\subsection{Mask, Shield and Onboard Calibration Source}


The mask consists of 12 identical stacked layers of 0.3 mm thick
Tungsten sheets, each of which has an identical pattern of a 2 \x 2
uniformaly redundant array (URA) chemically etched out in a 64 \x 64
square pixel configuration.  In theory, URAs allow a perfect imaging response
for a point source without coding noise.  This particular URA pattern
used for \pe1 is generated by perfect binary arrays
\cite{Busboom98}, which allow a square mask pattern ideal for our detector
geometry (see Fig.~\ref{hangerdetimg}). The mask pixel pitch is a 4.7 mm with 0.5 mm grid to support
isolated opaque elements. Therefore, the size of open elements
is 4.2 \x 4.2 mm\sS{2} and  the total open fraction is 40\% for normal
incidence (the open hole fraction is 48\%).  The mask, mounted on a
support frame, is located at 90 cm
above the detector plane, and the URA pattern covers 30 \x 30 cm\sS{2}
(each mask sheet spans 36 \x 36 cm\sS{2} for mounting), allowing
a 10\Deg \x 10\Deg fully coded field of view (FoV) or 
20\Deg \x 20\Deg for 50\% coding fraction.

\begin{table}
\small
\caption{Science Data Size and Rate}
\begin{tabular*}{0.47\textwidth}{r@{\extracolsep{\fill}}l}
\hline\hline
Parameters					& Values 	\\
\hline
Event header size\sS{a} + checksum		& 48 Bytes	\\
16 samplings per pixel				& 32 Bytes	\\
Raw event size per $N$ pixel readout		& $48+32N$	\\
Compressed event size per $N$ pixel readout	& $48+2N $	\\
\hline
In normal readout mode,			\\
Average number of pixels for readout		& 9.12 pixels	\\
Average size of raw event data 			& 339.8 Bytes	\\
Average size of compressed event data		& 66.2 Bytes	\\
\hline
Measured total event count rate 		& $\sim$ 750 cps\sS{b}	\\
Raw data rate					& 2.2 Mbps	\\
Compressed data rate				& 530 kbps 	\\
\hline
\end{tabular*}
(a) This does not include the packet header added during the
transfer between the FCB and the flight computer. 
(b) At 39 km float altitude, and dominated by cosmic-ray induced
events before shied rejection. X-ray background events (30--200 keV)
are $\sim$ 250 cps; see \S\ref{s:counting}.
\label{data}
\end{table}

Graded passive shields made of Pb/Sn/Cu were used to limit the
X-rays incident through the side of the telescope. These side shields are
divided into two sections (Fig.~\ref{pe1}). The upper section sits
on top outside of the pressure vessel (PV), mounted all around the
mask support tower above the top cover of the PV to the
mask. The lower section is mounted onto a similar support tower inside the
PV and covers all four sides of the detector plane above the rear active
shield to the bottom side of the top cover of the PV.  A part
of the side shields was not completely covered with Sn/Cu, and so the
lead sheets were partly directly exposed to the detector plane so that
fluorescent X-ray $K$\Ss{\alpha} and $K$\Ss{\beta} lines from the lead were observed in the spectrum (see
\S\ref{s:spectra}). The thickness of the side shields are 2.6/0.7/0.4 mm (Pb/Sn/Cu)
for the lower section, and 2.6/0.3/0.3 mm for the upper section.

In order to stop the albedo atmospheric X-rays and tag charged particle
induced background events, an active shield of a 2 cm thick 26 \x 26 cm\sS{2}
CsI(Na) scintillator crystal from
Scintitech\footnote{\url{http://www.scintitech.com}} is mounted about 10 cm
below the detector plane.  The CsI scintillator is larger than the active
area of the detector plane, extending out to the side shields in order
to reduce the direct open path to the detector plane from outside.
However, the current configuration leaves about 4 cm gap between the
rear and side shields, allowing a high background rate during
the flight (see \S\ref{s:counting}).
The optical scintillation photons are collected by two photomultiplier
tubes (PMTs,
Hamamatsu\footnote{\url{http://www.hamamastu.com}}
R1924A) on two opposite sides of the CsI crystal through edge-readout
wave shifters.  Preamps and subsequent electronics in a VME crate
provide a pulse for tagging shield events. A PC-104 based computer
(separate from the flight computer below) controls the VME electronics.
The FPGA on the FCB monitors the two discriminator signals from the active
shields, and if they occur within 2 $\mu$sec of the X-ray triggers, the
X-ray event is tagged.


In order to monitor the gain variations of the detector, a \sS{241}Am
calibration source (AmCal) was assembled and mounted. The AmCal source
consists of a 200 nCi \sS{241}Am doped in a plastic scintillator (Isoptope Products
Laboratories\footnote{\url{http://www.isotopeproducts.com/}}), which is
coupled to a PMT (Hamamatsu R7400U). The scintillation light induced from
the $\alpha$ particle coincident with a 60 keV X-ray is collected by the
PMT and amplified through a separate channel in the shield VME electronics
for tagging the Am events.  
In addition to the onboard calibration source, we can command all of
the 64 RadNET ASICs to inject on-chip pulses into every pixel at a
configurable frequency, and we read them out in the same way as 
normal X-ray events. This monitors the dead time and the electronics
gain variation of the readout system throughout the flight.

The fully assembled detector plane (Fig.~\ref{pe1}c) is mounted on an
Al frame and surrounded by a thin copper box on the four sides and the
rear side, which forms a detector shield enclosure.  The top cover of
the enclosure box is a 150 $\mu$m thick Al sheet.  Two fans are mounted
on the side of the enclosure and another two on the bottom for temperature
control through air (N\Ss{2}) circulation through the PV. The fan power is electrically isolated
from the power for the detector plane to minimize the electronics noise.


\subsection{Flight Computer, Data Rate and Transfer} \label{s:fc}

The flight computer is a 2 GHz Intel dual core compact PC (AEC-6920 by
AAEON Technology\footnote{\url{http://www.aaeon.com}}) with a 128 GB
solid state disk designed to operate in extreme environments.  The flight
computer collects the science and HK data from the FCB through Ethernet links,
stores them on the onboard hard disk, and sends a compressed version
of the data to the transmitter (see below).  It also issues to the FCB
the commands uploaded through a RS422 port from the gondola computer \cite{Allen11}.
In order to handle these tasks, the flight computer runs three basic
programs  -- a commmand listener (CL) program, a data acquistion (DAQ)
program, and a transmitter handler (TH) program under a linux (Fedora 8)
operating system.  A cron job based watchdog script monitors the health
of these three programs, and terminates or relaunches them automatically
if needed.

\begin{figure} \begin{center}
\includegraphics*[width=0.47\textwidth]{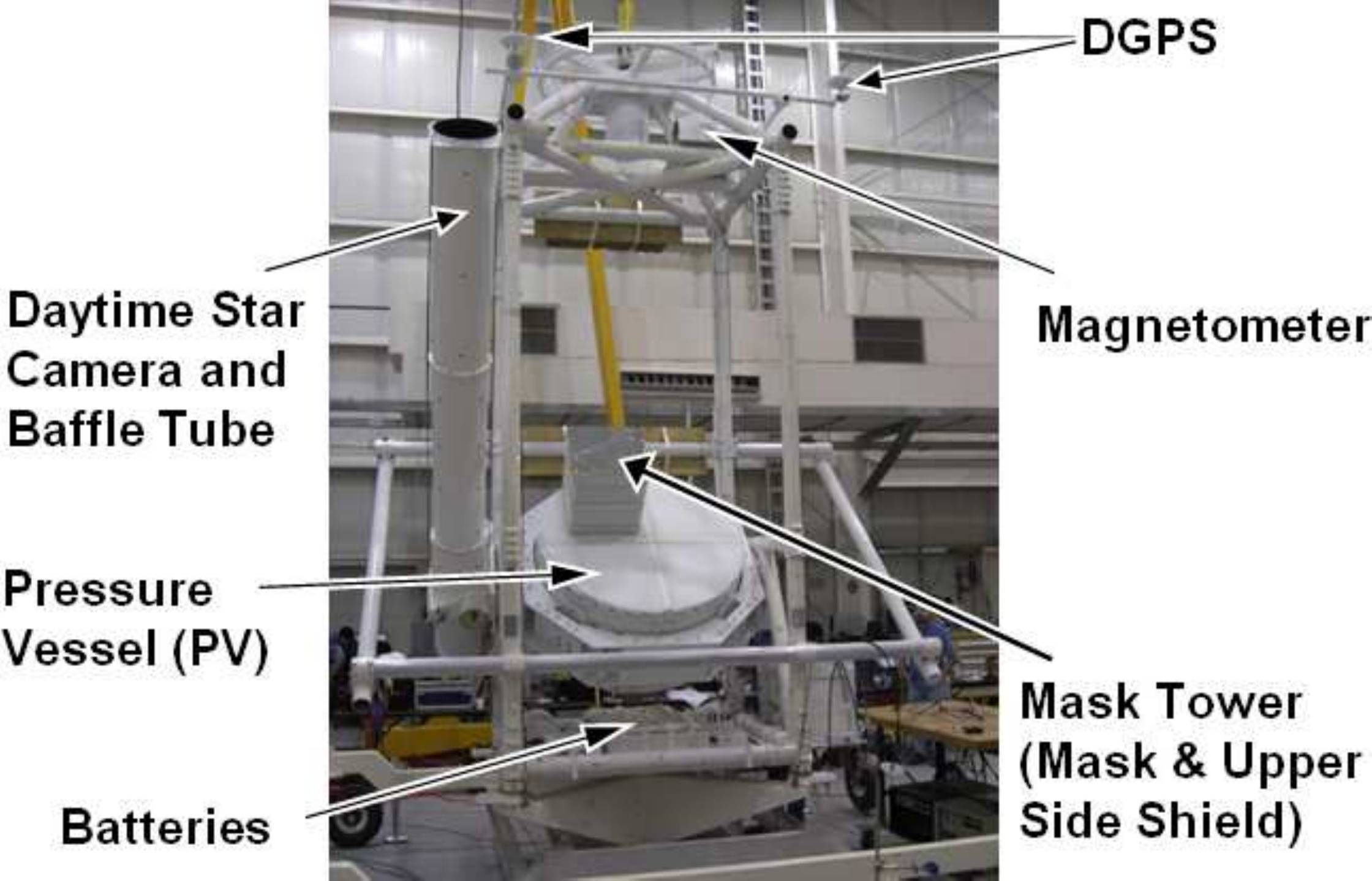}
\caption{The overview of the payload and gondola.}
\label{payload}
\end{center}
\end{figure}

Table 2 summarizes the data size and measured rate (see also Fig.~\ref{timeline}).
The raw data rate is about 2200 kbps for 750 cps.   The DAQ program
saves the entire raw data stream on the onboard solid state hard disk.
In order to meet the telemetry limit (400 kbps), the DAQ program
also compresses the data by calculating a single pulse height value from
16 samples of the pulse
profile (see Fig.~2 in \cite{Hong09}), and sends only the resulting
pulse height instead of the full 16 samples to the TH program. The
corresponding compressed data rate is about 530 kbps, which is about
24\% of the raw data rate instead of 8\% (1/16th) due to the overhead
of the event header. The compressed data rate can still exceed the
telemetry rate.  Therefore, the DAQ program monitors the data rate,
and drops a fraction of events randomly (only for telemetry) to meet
the telemetry bandwidth so that the full data stream for a given event
can be transmitted to the ground during the flight. The data reduction
fraction for telemetry is continuously adjusted during the flight to
keep the compressed data rate below 300 kbps, which allows a 100 kbps
margin for the packet header, the HK data or possible variations of
the telemetry limit.  For debugging purpose, the DAQ program can also
send the raw data without compression at request.  Transmission modes
(raw or compressed) and the data reduction fraction for the telemetry
can be controlled by the commands uploaded from the ground.

In addition to the science data, the HK data and the command echo
are merged and transmitted to the ground without any compression 
since their rate is relatively small.
There are four major HK data streams: the FCB HK data
reporting the various input voltages and currents including the HV
PS and the temperature of the FCB, the shield HK data reporting the
active shield and AmCal count rate and status of the HV PS for PMTs,
the DAQ HK data reporting the status of the flight computer such as free
hard disk space, and the Power HK data reporting the power
input voltages and the currents coming into the PV and temperatures and
pressures of a few points inside the PV.
On the ground, we
monitor and save the transmitted data, and issue commmands
accordingly.

\begin{figure*}[t] \begin{center}
\footnotesize
\includegraphics*[width=0.90\textwidth]{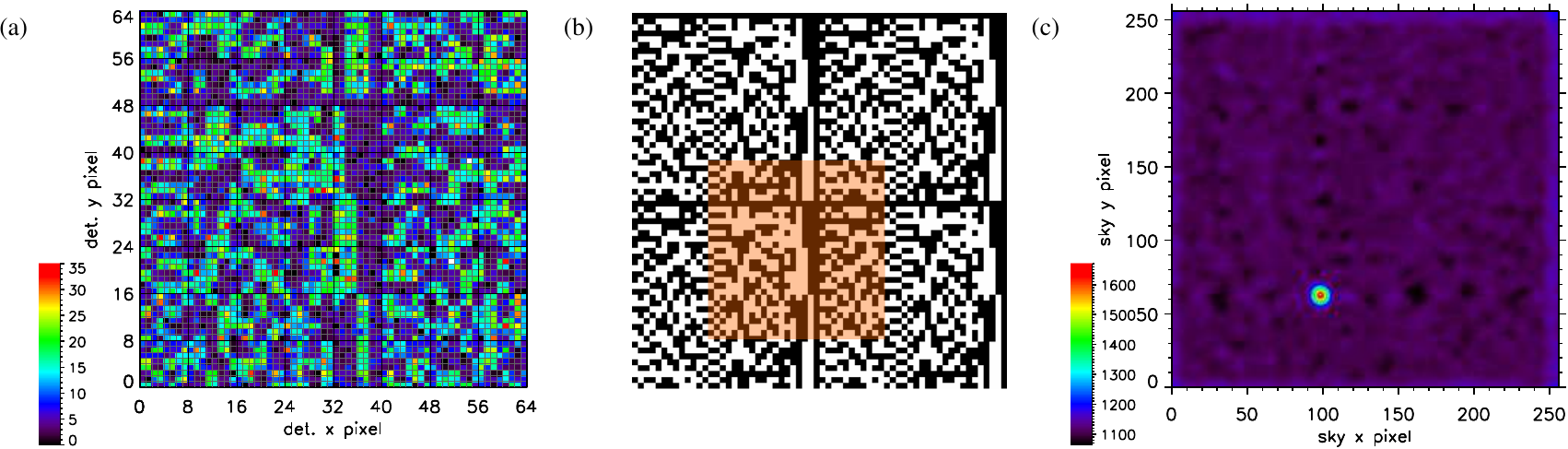}
\caption{(a) The detector count map generated by a 15 min irradiation of
an external \sS{57}Co source at 6.5 m away from the telescope.
(b) The 2 \x 2 URA mask pattern: the (orange) shaded
section of the mask pattern indiciates the coding area of this
particular configuration. (c) The reconstructed sky image in the fully coded FoV using
5.35 mm magnified mask pixel pitch. }
\label{hangerdetimg}
\end{center}
\end{figure*}

\section{Payload Integration and Test} \label{s:it}
The detailed description of the payload integration is
found in \cite{Allen10, Allen11}. Here we briefly
summarize the integration and test of the detector plane.

The CZT detector plane for the \pe1 telescope
was fully assembled (Fig.~\ref{pe1}c) at Harvard University on
2009 August 31.  In September, 2009, the detector plane was mounted inside
the PV along with the side passive shields, the CsI rear active shield,
the AmCal source and the electronics.  Then, the \pe1 telescope payload
- the PV and the mask tower - was integrated into the Harvard gondola
now including  a new pointing and aspect control system consisting of
a daytime star camera, differential global positioning system (DGPS)
units, and a magnetometer (Fig.~\ref{payload}) \cite{Allen10, Allen11}. 

After the payload integration, we irradiated
the X-ray telescope using
a $\sim$ 5 mCi \sS{57}Co radioactive source. The source was
positioned about 6.5 m away from the mask within
a few degrees of the X-ray telescope axis.  
Fig.~\ref{hangerdetimg}a shows the detector count
map of the source in the 80 -- 150 keV band from about 15 min of
the irradiation of the \sS{57}Co source. For illustration,
the 2 \x 2 URA mask pattern is shown 
in Fig.~\ref{hangerdetimg}b with
a section highlighting the matched pattern observed in the
detector image.
Fig.~\ref{hangerdetimg}c shows the reconstructed sky image of the fully
coded FoV using a balanced cross-correlation (e.g.~see \cite{Roques87})
between the mask pattern
and the detector image and accounting for magnification of the shadow
pattern due to the finite source distance.  The measured $S/N$ ($\sim$
80) of the source is a bit lower than the Poisson-noise based $S/N$
because of the coding noise introduced by incomplete sampling (91\%)
of a full cycle of the URA pattern due to magnification.

For energy calibration, we use the pre-flight radiation data 
using the \sS{57}Co source.
For a given pixel at a given trigger
capacitor, we identify the raw ADC values of three known energies  -
zero keV from the reference pixel measurements, 122 keV from the
\sS{57}Co source and 650 keV from the internal pulser inputs - 
by a fit using a gaussian function to the raw spectra.  The conversion
from other raw ADC values to energies uses a 2nd order polynominal fit
between the raw ADC values and X-ray energies of the three known values
\cite{Hong06}.  After this relatively quick calibration, the detector
shows about 4 keV FWHM at 60 KeV lines and 10 keV at 120 keV.  The latter
indicates that there is room for improvement in the calibration (see
\cite{Hong06,Hong09,Allen11}): e.g.~the pulser is a less reliable
energy calibrator, and the charge split or depth
correction is not performed.

\begin{figure*}[t] \begin{center}
\includegraphics*[width=0.85\textwidth]{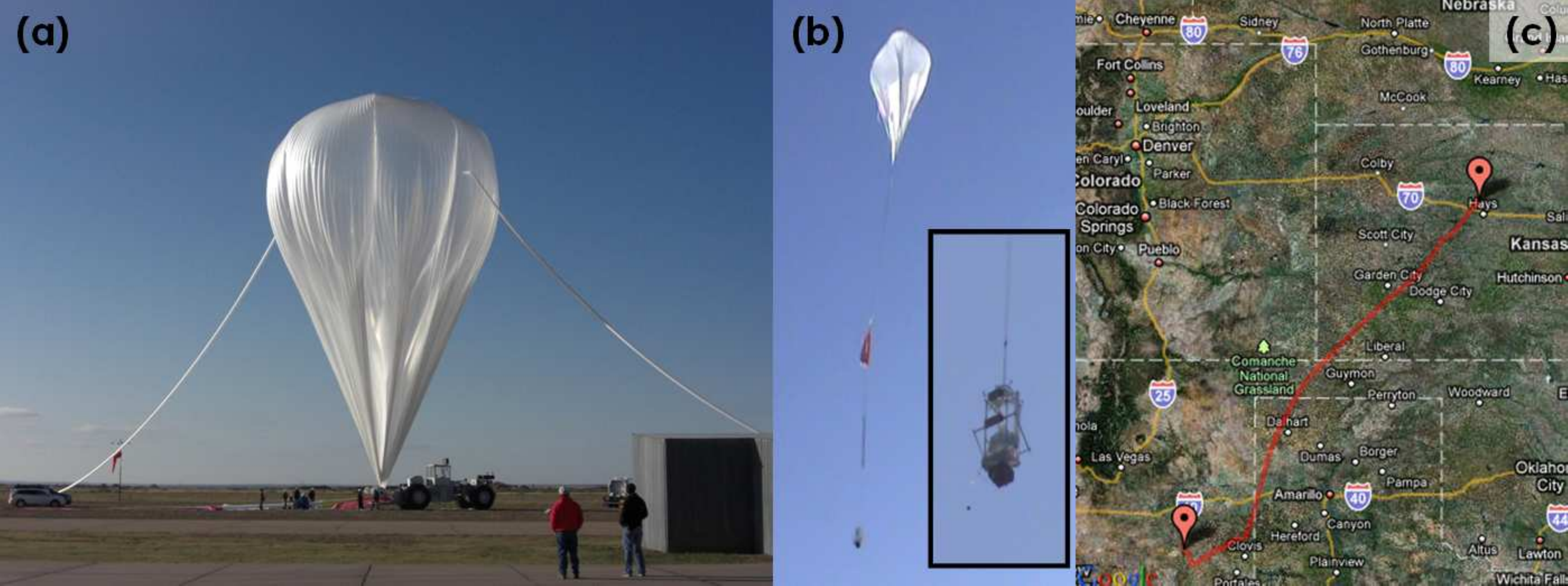}
\caption{The inflated balloon "bubble" (top $\sim$ 1/4 of balloon) right before the launch (a), the liftoff (b) 
and the flight path on a Google map (from Ft.~Sumner, New Mexico to
Hayes, Kansas) (c).
The inset in (b) shows the payload during the liftoff.}
\label{launch}
\end{center}
\end{figure*}

The tagging efficiency of the AmCal events in the CZT detector is measured
about 74\% (see \S\ref{s:counting}), and does not increase
with the coincidence window as long as the window is
longer than 2 $\mu$sec. Therefore, the other 26\% AmCal events are
suspected to fail to generate a tagging signal all together probably
due to insufficient light collection in the PMT from the 5 MeV $\alpha$
particle of \sS{241}Am decays near the edge of the scintillator.
The tagging efficiency of high energy charged particles by the CsI
shield is estimated to be around 50\% on the ground, based on the high
energy events induced by atmospheric muons, which are recorded near the
upper limit ($\sim$ 1 MeV) of energy deposit for a pixel.  This is
roughly consistent with the solid angle coverage of the rear shield.
This tagging efficiency is likely energy dependent.

\section{Flight Performance}\label{s:flight}

\subsection{Flight Overview}

After 6 previous launch attempts, early in the morning on Oct 9, 2009,
all the systems were checked twice with an external power supply for
flight, and then before balloon inflation began, they were switched
to the onboard battery, waiting for launch.  The procedures for the
initialization sequence of the detector are, in order: powering up the
system, activating the internal ASIC pulsers, disabling the known hot
pixels, applying the HV bias on the CZT crystals at $-$600 V (gradually
ramped up over $\sim$ 10 min), and setting the DCU thresholds gradually
down to $\sim$ 30 keV over $\sim$ 15 min.  Each DCU has its own unique
threshold setting but all  64 pixels in a DCU share a common threshold.
The complete sequence takes about 30 min and each procedure can be
either manually or autonomously executed.  The internal ASIC pulsers
were run in all of the 64 DCUs at 0.3 Hz  throughout the flight for a
dead time calibration.  The active shield system is also initialized
right after power-up.


The payload was launched at 14:30 UT by a 40 million m\sS{3} high
altitude balloon from the NASA CSBF at Ft. Sumner, New
Mexico. Fig.~\ref{launch} shows (a) the inflated balloon bubble (top $\sim$
1/4 of the balloon) right before the launch, (b) during the liftoff and (c) 
the flight path on a Google map.
Fig.~\ref{timeline} shows the aspect and pointing history of the system
during the flight and the raw count rate history of several key event types
recorded in the detector system.  The payload reached the float altitude
of 40 km at 17:00 UT (blue in
Fig.~\ref{timeline}a). The flight duration was only about 7.5 hr with
5.5 hr at float altitude due to strong winds ($\sim$60 knots) and it was
terminated near Hayes, Kansas, due to both the $\sim$ 500 km telemetry
limit and the NASA safety regulations that prohibit flights over densely
populated area.  The landing was relatively harsh due to high surface
winds, and the gondola frame sustained some damage, but the payload was
recovered without any damage.

When the payload reached the float altitude, the Crab pulsar, one of
a few sources bright enough for detection in a short observation for a hard X-ray telescope of
this moderate size, had just set. Until Cyg X-1 rose
about 4 hr later, we succesfully performed a number of system checks such
as shield threshold tests (18:00 UT), while attempting tracking 3C 273.
From 21:00 UT, we observed Cyg X-1 until the termination of the flight.
After an hour observation of Cyg X-1, we reset the power of the detector
system at 22:00 UT for a test. The system came back up without a glitch.
After the power reset and the initialization sequence, we had less than
a 10 min of fine-pointed observation of Cyg X-1 before we were requested to terminate
the flight.

The X-ray telescope performed excellently throughout the flight, but
the aspect and pointing system suffered a few major issues:  until in
the middle of the Cyg X-1 observation, the telescope could not track
a target properly.  First, we lost the DGPS units, possibly due to
insufficient thermal insulation, at around 15:46 UT during the ascent.
Second, a malfunction of the star camera software disabled the live
feedback to the pointing system and forced us to rely on the magnetometer
and the elevation axis for target tracking. 
The absolute error of the
raw azimuth values reported by the magnetometer can be as large as 10\Deg
without calibration, depending on the orientation.  
Third, the elevation axis
failed to follow the target properly until stablized later in
the flight (21:55 UT), likely due to insufficient 
torque for the motors.

Despite these issues, Cyg X-1 was within the partially coded FoV above
60\% coding fraction and it was mostly $\lesssim$ 2--3\Deg off in the
elevation axis and 6--7\Deg off in azimuth from the
X-ray axis (see Fig.~\ref{timeline_cygx1}).  We calibrated the raw azimuth
values using those reported by the DGPS units during the accent before
they failed when the payload was freely spinning (14:30 - 15:46 UT;
black line in Fig~\ref{timeline}a). This calibration reduces the absolute
error of the azimuth down to $\sim$ 1 -- 2\Deg.
Fortunately, a set of the
sky images captured by the star camera were later identified and used
to verify and correct the aspect and pointing history derived by the
magnetometer and the elevation axis. The time stamps of these images
are marked by (green)
`X's at the right-most side in the top panel of Fig.~\ref{timeline}.

\begin{figure}[t] \begin{center}
\footnotesize
\includegraphics*[width=0.47\textwidth]{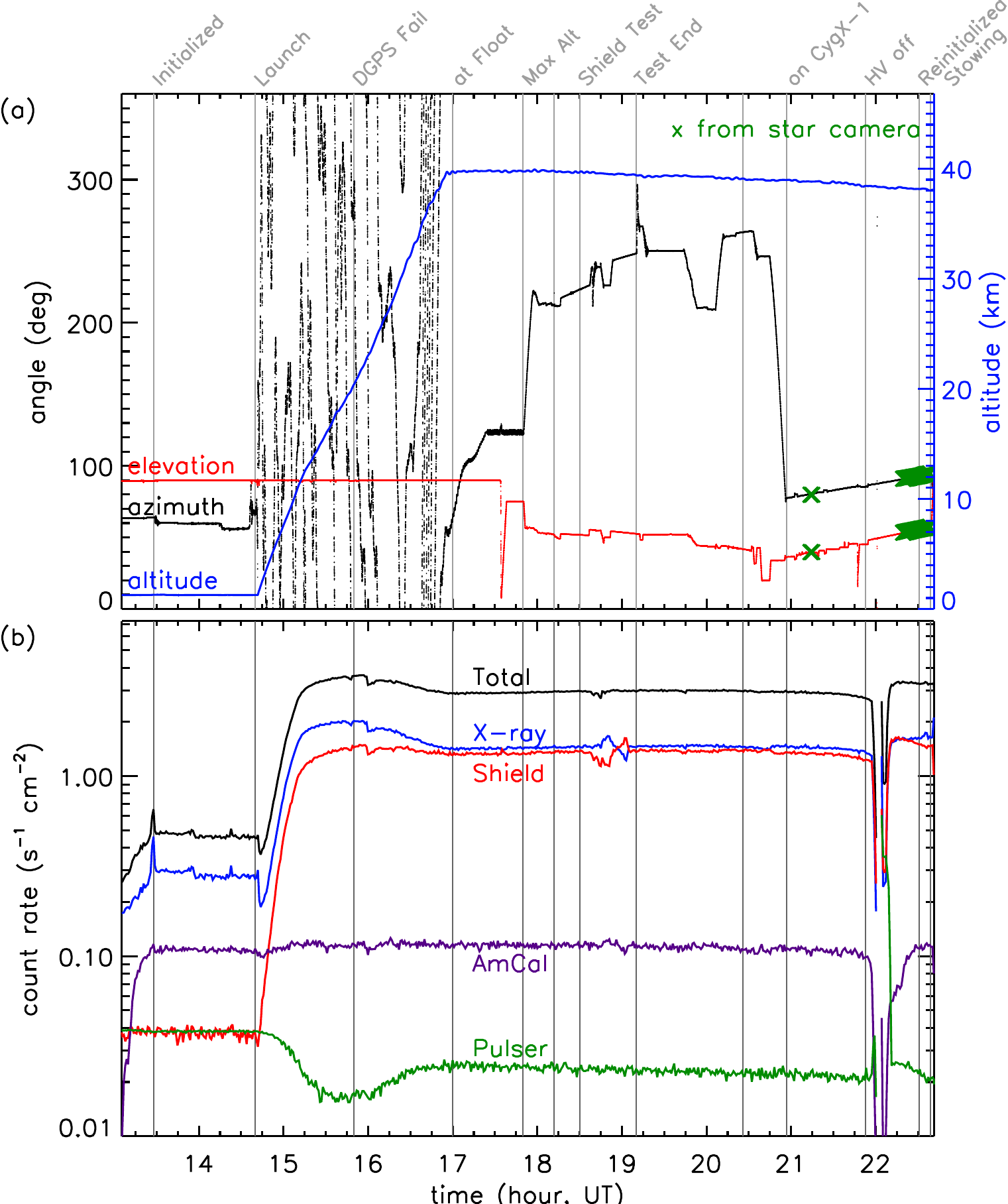}
\caption{(a) The aspect and pointing history of the flight. The
azimuth angles (black) are based on
the DGPS-calibrated magnetometer, and the elevation angles (red) are
based on the elevation axis.  The altitude (blue) is recorded by 
external pressure gauges. The (green) `X's mark the azimuth and
elevation values drived by a set of sky images captured by the star
camera. The top axis labels the key events of the
flight.  (b) The raw detector count rate history of the flight. From
top, the total event rate (solid black), the X-ray event rate
(blue), the shield tagged event rate
(red), the AmCal tagged event rate (solid
blue), and the pulser event rate (green). The dip in pulser rate at
15--16 hr is due to increased dead time at the high count rates
encountered in the Pfotzer maximum in balloon ascent.
}
\label{timeline}
\end{center}
\end{figure}

\subsection{Event Rate} \label{s:counting} \label{s:class}

Fig.~\ref{timeline}b shows the raw count rate history of the detector
system during the flight. The total event rate (black) is the total
trigger rate in the detector system, which is dominated by the cosmic ray
induced interactions.  The shield tagged event rate is shown in red, and
the X-ray event rate (blue) is from the events without a shield tag. The
AmCal rate (purple) is the event rate with an AmCal tag.  The pulser rate
(green) is the internal ASIC pulser rate.  Note that each raw event is
recorded in a unit of a DCU pair, counting a multi-pixel trigger event as
one event but counting a multi-DCU-pair trigger event as multiple events.
The latter means that the 0.3 Hz pulser injected over the entire detector
plane should be recorded with rate
0.0394 counts s\sS{-1} cm\sS{-2}
when there is no dead time\footnote{0.0394 = 0.3 Hz \x 32 DCU pairs
/ 243.4 cm\sS{2}, where the detector area is 1.95 \x 1.95 cm\sS{2} \x
64 DCUs.}.


The absolute timing accuracy of events is in an order of 0.5 $\mu$s, which
is determined by the sampling frequency of the RadNET ASIC.
We save the
time tag of each event with a 0.1 ms resolution.  A typical readout time
for an event with a 10 pixel readout is about 5 ms, which leads to a
dead time for event readout in a given DCU pair, but not for the rest of
the DCU pairs since each DCU pair operates independently.  
The analysis show that almost all of the X-ray events
with multi-pixel triggers in a DCU pair are truly simultaneous (not a
random coincidence of multiple events), whereas about 9\% or less of the
X-ray events with multi-DCU triggers are truly simultaneous.  In the
case of multi-pixel triggers, almost all of the X-ray events are four
pixel triggers or less.  Therefore, for imaging and spectral analysis
of Cyg X-1 and the AmCal events, we mainly focus on the low energy
X-ray events with four pixel triggers or less, and we consider all the
low energy X-ray events with multi-DCU-pair triggers as independent events.
For the charged particle induced events with a shield tag, about 30\%
triggered multiple DCU pairs. Many of them are truly simultaneous,
generating large tracks (Fig.~\ref{track}), which illustrates a
potential of large area CZT imaging detectors as cosmic ray detectors.

\begin{figure*}[t] \begin{center}
\includegraphics*[width=0.90\textwidth]{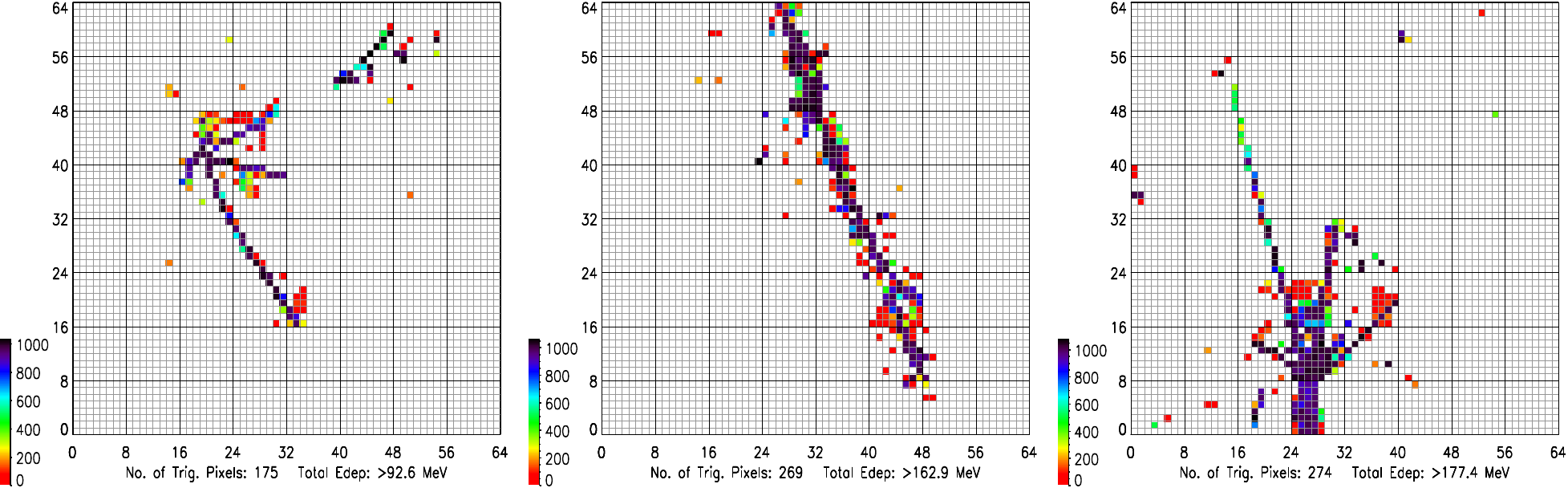}
\caption{Examples of multiple DCU trigger events 
induced by 
cosmic ray interactions during the flight.  The
number of the simultaneously triggered pixels and the total energy
deposited by the events are labeled. The color scale shows the energy
deposit in each pixel in keV. Since the energy deposit is
saturated at $\sim$ 1 MeV, the recorded
total energy deposit is a lower limit.
} \label{track}
\end{center}
\end{figure*}

The raw X-ray event rate in Fig.~\ref{timeline}b contain AmCal events due to
missing tags and is not corrected for dead time. To extract the proper
X-ray event rate and spectra, we use the following relations.
\begin{eqnarray}
	r\Ss{X} & = & f\Ss{Live} (R\Ss{Am} + R\Ss{X}) - r\Ss{Am} \\
	r\Ss{Am}& = & f\Ss{Live} (f\Ss{AT} R\Ss{Am}  + f\Ss{RT} R\Ss{X}) 
\end{eqnarray}
where $f\Ss{Live}$, $f\Ss{RT}$, $f\Ss{AT}$ are the live time fraction,
the random coincidence fraction by AmCal tag and the AmCal tagging efficiency
respectively, and
$r\Ss{X}$ and $r\Ss{Am}$ are, respectively, the measured raw count rates
of the X-ray events (no shield or AmCal tag) and the AmCal events (an
AmCal tag, but no shield tag);
$R\Ss{X}$ and $R\Ss{Am}$ are the corresponding true count rates
(i.e. the incident rate if the detection efficiency is 100\%). 
Fig.~\ref{ctimeline} shows these corrected rates ($R\Ss{X}$ and
$R\Ss{Am}$) for various cases and Fig.~\ref{specov}
for the energy spectra
before and after correction. 

The live time fraction ($f\Ss{Live}$) is estimated based on the
captured internal pulser fraction and random coincidence rate of AmCal
tagging
($f\Ss{RT}$) is estimated based on hard X-ray events in the 80--200 keV band
where all the AmCal tagged events are due to random coincidences.
(Fig.~\ref{ctimeline}a).  Pulser
events are easy to identify due to simultanoues injection of 
650 keV equivalent signals into all 128 pixels of each DCU pair,
which produce a narrow line around 83 MeV (for summed pixels) in the pulse height histogram.
(e.g. green in Fig.~\ref{specov}c).  The pulser recovery fraction
($=f\Ss{Live}$) is about 97\% on the ground,
and it dropped down to 40\% during the ascent due to
a large increase of background, and then stayed around 55--60\% during
the flight.\footnote{
The event
readout time in a DCU pair ranges from $\sim$ 5 to 25 ms, 
depending on the number of readout
pixels.  
For an event occurance rate of 1500 cps over the
full detector plane, which is equivalent to about 47 cps per a DCU pair, the
event capture fraction under an on-average $\sim$
15 ms event readout time 
would be about 50\% ($=\exp(-47 \times 1.5\times 10^{-2}$)).
Therefore,
the observed total event rate of $\sim$ 750 cps is consistent with the
$\sim$ 50\% live time fraction.}
Fig.~\ref{ctimeline}a (red) shows $f\Ss{RT}$ stayed fairly constant
throughout the flight. For the
rest of the calculation, we use the average value of $f\Ss{RT}$,
3.6\%.

We set $f\Ss{AT}$ so that the resulting energy histogram of the X-ray
events free of the AmCal events shows no positive or negative bump around
the 60 keV (black curve in
Fig.~\ref{specov}b).  We did this for both the flight spectra and the
prelaunch ground
spectra, and for $f\Ss{Live}$, we use the average value of each data set.
In order to have a smooth continuum at 60 keV, the allowed range for
$f\Ss{AT}$ is between 0.735 and 0.740; we therefore used 0.738 
for $f\Ss{AT}$ for the calculation.

Fig.~\ref{ctimeline}b shows the corrected total event (gray), X-ray
event (black) and AmCal (red) event rates. 
The total event rate (grey) reached about 9
counts s\sS{-1} cm\sS{-2} during the ascent and dropped to 4.5 counts
s\sS{-1} cm\sS{-2} at the float altitude and gradually increased to
5.4 counts s\sS{-1} cm\sS{-2} later in the flight. 
The X-ray event rate (black) stayed about 2.3 -- 2.7 counts  s\sS{-1}
cm\sS{-2} at the float altitude. 
Since the trigger rate in the CsI shield ($\sim$ 100 keV threshold),
which likely represents the charged particle induced event rate, (not shown in the figure)
was 2.1 -- 2.7 counts s\sS{-1} cm\sS{-2},
the combined rate of the X-ray and charged particle induced
events makes up the observed total count rate of 5.4 counts s\sS{-1}
cm\sS{-2}.  However, the X-ray event rate still contains a large
fraction of charged particle induced
events, considering the relatively low shield tagging efficiency
($\sim$50\%, \S\ref{s:it}), a higher threshold of the CsI shield, and
secondary background due to interactions in the surrounding side
shield and mounting structures.  To resolve the precise composition of the X-ray
background rate, detailed Monte-Carlo simulations of X-ray and
charged particle transport in the telescope geometry are needed.

The low energy X-ray events (30--200 keV, blue) stayed
at around 0.86 -- 1.1 counts s\sS{-1} cm\sS{-2}.
If we assume a shield tagging efficiency of 50\% (\S\ref{s:it}),
the low energy X-ray event rate was about 0.5 -- 0.7 counts s\sS{-1}
cm\sS{-2} (purple).
Assuming an albedo X-ray flux of $13.9 E^{-1.81}$ 
ph cm\sS{-2} s\sS{-1} keV\sS{-1} at 3.5 g cm\sS{-2} \cite{Gehrels85}
and an aperture FoV of 0.1 sr (FWHM) with a 40\% open fraction,
we expect the aperture X-ray background rate of
$\sim$ 0.04 counts s\sS{-1} cm\sS{-2} in the same band.  In addition, the
gap between the rear and side shields allows a projection-corrected FoV of
0.5 sr, which can contribute $\sim$ 0.45 counts s\sS{-1} cm\sS{-2}.
We also estimate a total flux of $\sim$ 0.1 counts
s\sS{-1} cm\sS{-2} in the Pb fluorescent X-rays.
The combined aperture and leakage X-ray flux in the 30-200 keV band is
estimated to be $\sim$ 0.6 counts s\sS{-1} cm\sS{-2}, which is consistent
with the observed X-ray event rate.  This estimate lacks some details such
as attenuation by surrounding structures for X-rays coming below,
the contribution from the internal
background or an increase of the active area due to gaps between CZTs
but their effects are rather small or their contributions cancel each
other to some extent.

The corrected AmCal rate ($R\Ss{Am}$, red) stayed 
relatively constant at 0.128 $\pm$ 0.006 counts s\sS{-1}
cm\sS{-2}, indicating our analysis works well.  Based on the known source
strength of the AmCal sources with
$\sim$ 2\% attenuation from the detector enclosure window,
etc, one can calculate the overall detection efficiency at 60 keV to be 0.86
$\pm$ 0.04; the $\sim$ 10\% loss of QE is due to a variety of effects
(dead surface layer, etc) and will be reported for several energies in \cite{Allen11}. 
The photo-peak efficiency in the 55 -- 65 keV range is
0.54 $\pm$ 0.02. The error values are only of statistical origin,
and the simplified geometric assumption of a point source can add
an additional 5\% systematic error.

\begin{figure}[t] \begin{center}
\footnotesize
\includegraphics*[width=0.47\textwidth]{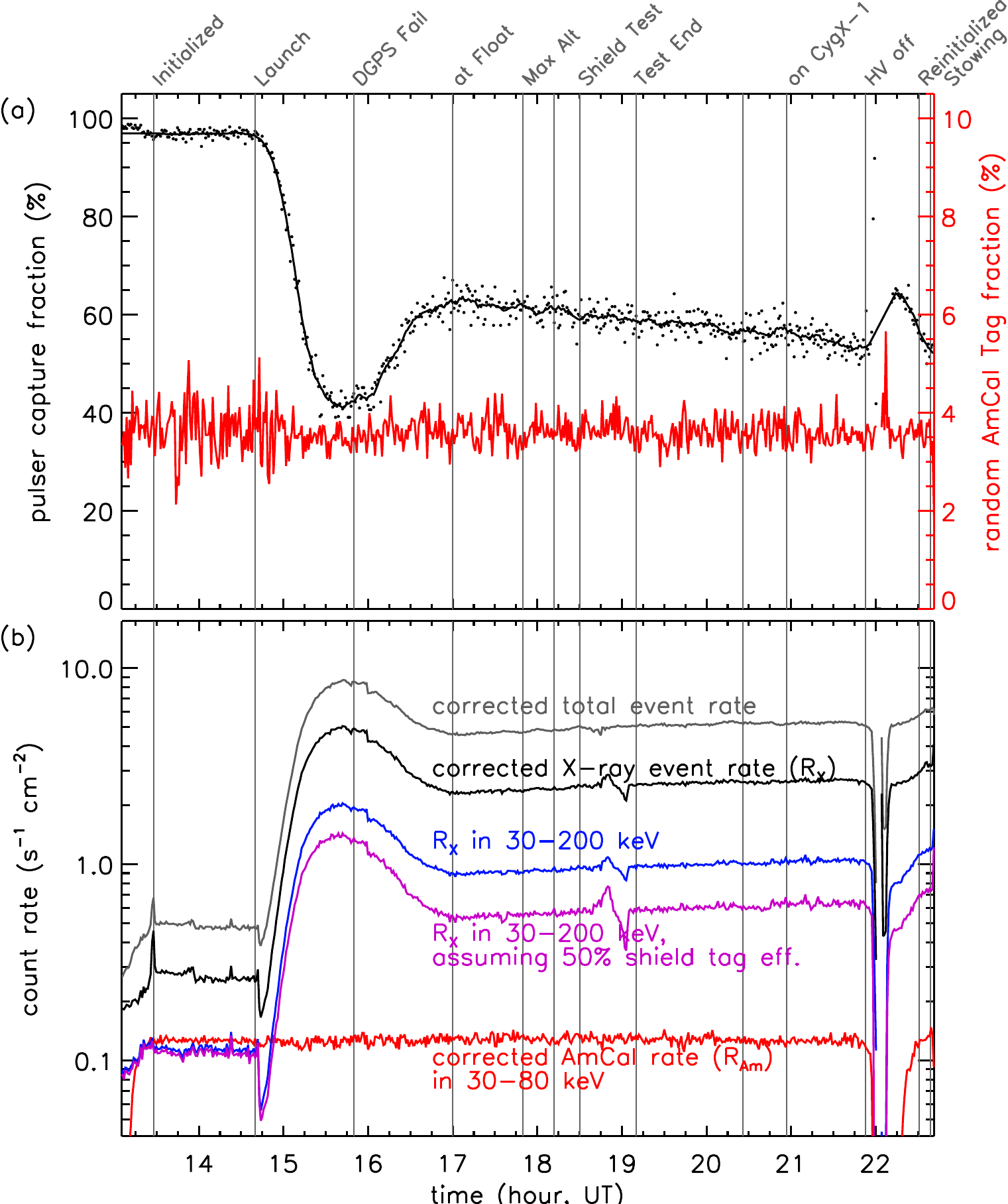}
\caption{(a) The pulser capture fraction (= the live time fraction,
black) and random tagging
fraction of the AmCal source events (red). (b) The count rate
history of the corrected data by Eq.~(1) \& (2). Note the
correction fixes only the untagged AmCal event rate, so that
the X-ray event rate still includes contributions from the charged
particle induced events.
}
\label{ctimeline}
\end{center}
\end{figure}

\subsection{Detector Spectra} \label{s:spectra}

Fig.~\ref{specov} shows the energy histograms of various event types
before and after correction by Eq.~1 and 2 in \S\ref{s:class}.
The top panels show the low energy section of the histogram in photon
flux, and the bottom panels show the full range of the histogram in
energy flux unit.
Fig.~\ref{specov}d breaks up the X-ray events (no shield tag, no AmCal
tag) by the number of triggering pixels.
The flight data in the figure are from the events recorded between
17:50 and 21:53 UT, and the ground data between 13:35 and 14:40 UT. In
the case of the corrected AmCal spectra 
the ground data (not shown) agree with the flight data
(red in Fig.~\ref{specov}b \& c). 
Note that, after correction, the AmCal spectra lack
high energy events ($>$65 keV).

A 5 mm thick CZT detector is about 80\% transparent to 500--600 keV
photons, and the higher energy events in the spectra are mostly from
the multi-pixel triggering
events such as the charged particle induced events and the pulser events
at 83 MeV (green) (Fig.~\ref{specov}c).  The peaks at 75 and 85 keV
are due to the K$\alpha$ and K$\beta$ lines of the lead fluorencent X-rays
as aforementioned. A mild bump around 511 keV originating from electron
positron pair production is visible 
in the single pixel trigger event (blue) of
Fig.~\ref{specov}d.  The origin
of the peak at $\sim$ 1 MeV is not cosmic but is due to the upper limit of
the dynamic range of the ASIC preamp channel (i.e. the maximally allowed
energy deposit for a pixel), and the multi-pixel readout generates the
additional peaks at its multiples around 2 and 3 MeV.

\begin{figure*}[t] \begin{center}
\footnotesize
\includegraphics*[width=0.90\textwidth]{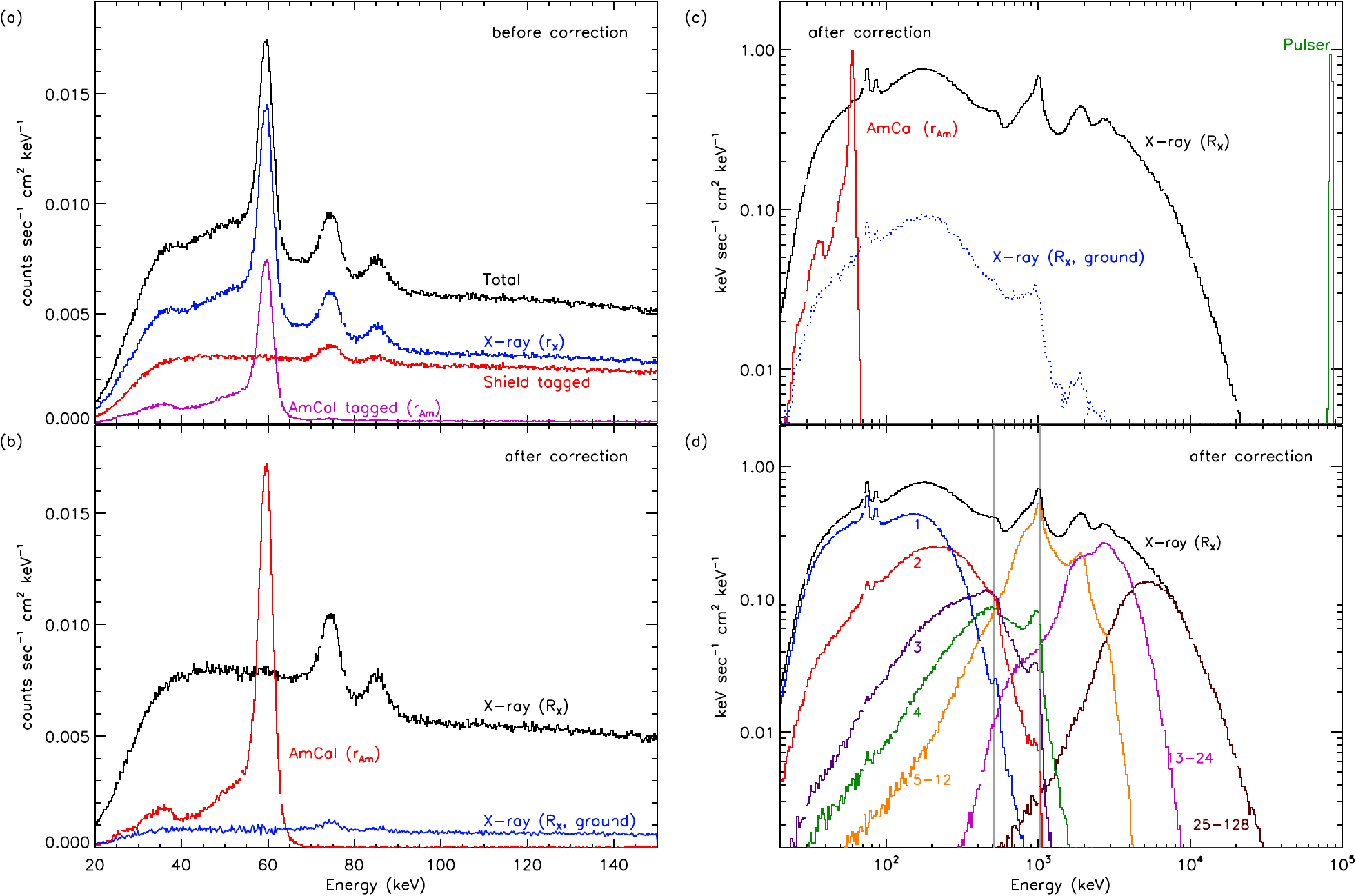}
\caption{(a) The energy-calibrated observed flight spectra
classified by shield and AmCal tags,
(b) \& (c) the corrected flight spectra
by Eq. 1 \& 2 in comparison with the corrected
ground X-ray spectra (blue), and
(d) the corrected flight
spectrum of the X-ray events and their composition by the (labeled) number of
triggering pixels.
The left panels show the spectra in the photon flux and the right
in the energy flux with a wider energy range.
In (a), the X-ray spectra ($r_X$) represent
events without an AmCal or shield tag, but 
they still contain the AmCal events, as indicated in a large peak at 60
keV. After correction in (b) and (c), the X-ray spectra ($R_X$)
do not have any contribution from the AmCal source.
In (c) the X-ray spectra of the
corrected flight data (black) are substantially brighter than those
of the ground data (blue),
but the AmCal and pulser spectra of the corrected flight data
are nearly identical to those of the ground data (not shown). 
The vertical lines in (d) mark
511 keV and the maximally allowed energy deposit
in a pixel ($\sim$ 1 MeV); higher energy peaks are due to 
2 or 3 pixel pileup. }
\label{specov}
\end{center}
\end{figure*}

\subsection{Temperature Dependent Gain Variation}

The detector system is mounted inside of the PV, which is
pressurized with dry N\Ss{2} and thermally controlled by a set of
heaters and fans.  The detector module experienced a $\sim$
10{\Deg}C temperature variation during the flight.
Fig.~\ref{amcal} shows the AmCal spectra summed over the entire detector
plane as a function of time in comparison with the temperature variation
(blue).  For easy comparison, the energy histogram is renormalized
at each time bin.  The figure shows  that the detector gain correlates with
the temperature 
($\Delta E$ (keV) $\approx 0.116 \Delta T$
({\Deg}C)). The temperature dependent gain change is likely due to the
capacitance variation of sampling capacitors in the ASIC, and the CZT
properties do not change observably over this temperature variation.
A similar trend is observable with the pulser run.  Unlike the gain,
the energy resolution does not vary much with the temperature.
For more information on the temperature dependent detector response of
the \pe1 detector, see \cite{Allen11}.

\section{Imaging Cyg X-1}

We observed Cyg X-1 for about 100 mins at the end of
the flight.  After an hour into the observation, which had large
pointing excursions (mostly in elevation; see Fig.~\ref{timeline_cygx1}a),
we performed a power cycle test of the detector system, which took about
30 min. Thus we have a total of an hour and ten minutes of usable
data with only the final 10 min having relatively stable ($\pm$
10\arcmin) pointing (see Fig.~\ref{timeline_cygx1}).  Here we decribe
the aspect calibration, image processing and detection
of Cyg X-1 as well as its measured energy spectrum.

\subsection{Image Reconstruction}

The event distributions in the detector observed during the flight are
substantially different from those measured on the ground.  For instance,
the dominance of charged particle interactions increases the fraction
of multi-pixel events, which more likely produce an artificial pattern
of event distribution.  In low energy X-ray events with $\le 4$ pixel
triggers, there are a few factors causing the non-uniformity
in the event distribution in addition to the usual pixel or crystal
dependent efficiency variation.
Fig.~\ref{bkgnd}a shows an example raw detector image of single pixel
trigger events. In order to visualize the non-uniformity more clearly, we
also show the event distribution folded onto a DCU pair (i.e. all 32 DCU
pairs are stacked), as illustrated in the small panel. In this example,
the first noticeable non-uniformity is the count enhancement of the edge
pixels due to gaps (900 $\mu$m for 1.95 cm \x 1.95 cm crystals) between
CZT crystals that provide additional surface area for accepting off-axis
X-ray events. In addition, the current sequential readout scheme performed in an
increasing order of pixel number makes the data of the last few pixels
more susceptible
to corruption.  In summary, non-uniformities in the detector image
have various origins: for some, the cause is obvious and for others,
not. For imaging, we
rely on background subtraction to eliminate the
non-uniformities regardless of their cause. We also explored an option of
using a synthetic background pattern for subtraction, which was generated
based on the observed non-uniformities. The synthetic background pattern
does not work as well, which is likely due to unknown and thus missing
components of the non-uniformities in the synthetic pattern.

\begin{figure*} \begin{center}
\footnotesize
\includegraphics*[width=0.95\textwidth]{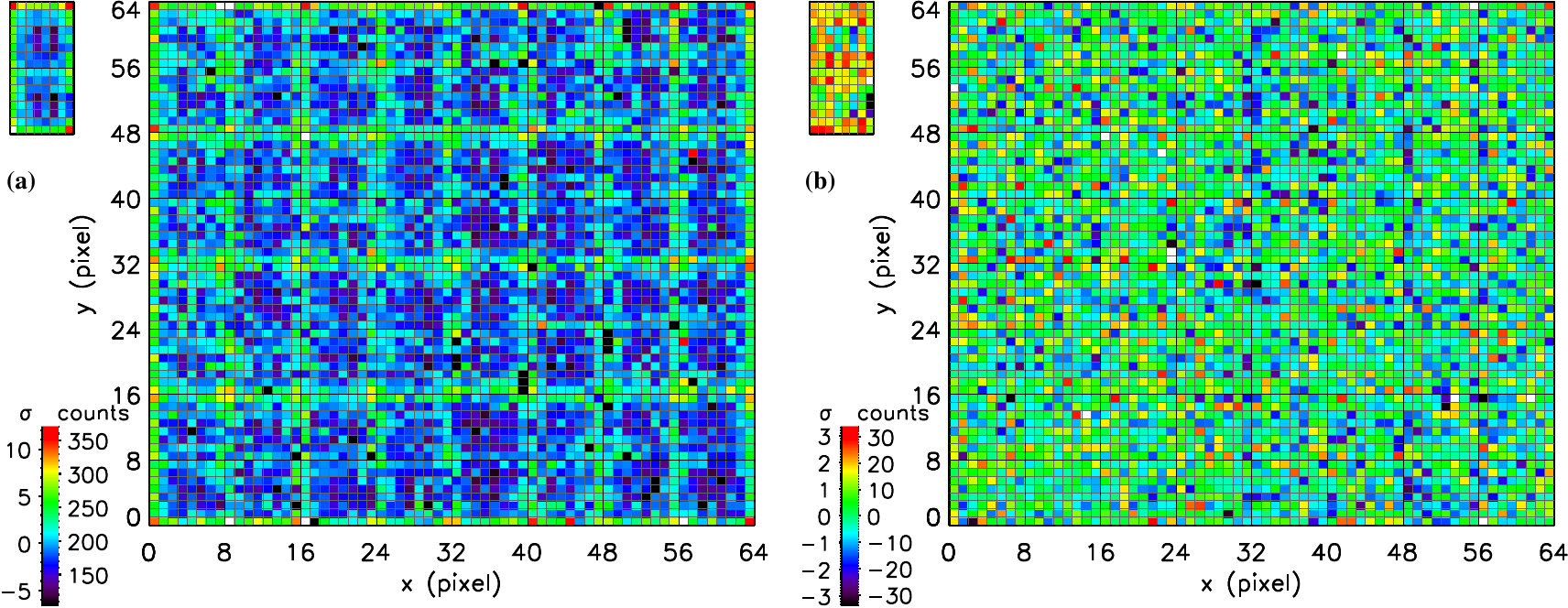}
\caption{Example detector plane images for 
single pixel trigger events in 30 -- 200 keV: (a) before background
subtraction, the count varies from pixel to pixel more than 10$\sigma$
above the average and (b) after background subtraction, the variation
is within $\sim$ 3$\sigma$. The DCU-pair folded event distribution
in the left top of each panel reveals the non-uniformity more clearly
before the background subtraction: e.g. an excess of edge pixel counts
and a deficit in a pattern that resembles the number '11'.
}
\label{bkgnd}
\end{center}
\end{figure*}

We generated a background detector plane image using the data set
between 18:30 and 20:58 UT, which is just before pointing on Cyg X-1
and so the closest in-flight background free
of bright X-ray sources.  Then we subtracted the scaled (for total
counts) background image from the detector plane images of the
observation.  Note the overall
background rate varies with altitude and elevation, but the pattern
of non-uniformities appears less sensitive to altitude or elevation as
demonstrated by the uniformity of the background subtracted
images below (Fig.~\ref{bkgnd}b).
Since the background dominates the source counts
($<5\%$), the scale factor used for background subtraction is simply the
ratio of the total count of the observation to the total count of the
background image, which makes the total counts of the subtracted image
zero.  The detector had 37 dead or disabled pixels (out of 4096 total),
which include two pixels disabled during the flight.  Even after disabling
these hot pixels, there are a few pixels (usually less than five) showing
about a factor of 5 or 10 larger count rate than the rest 
(8-10$\sigma$ above the average).  In a detector
plane image, we set five\footnote{The final results are not sensitive
to the exact choice of this number (from 5 to $\sim$ 10) as long as
the extreme outliers (8-10$\sigma$ above the average) are removed.} 
highest pixel counts to the next-highest value.

Fig.~\ref{bkgnd}b shows an example of the background subtracted detector
plane images using single pixel trigger events in the combined set from
the Cyg X-1 observation. The background subtraction
eliminates most of the outstanding non-uniformities.  For the Cyg X-1
observation, we generate such a background subtracted detector image
every 12 s from 20:57:38 to 21:53:02 UT before the system reset (the
power-cycle test) and from 22:31:08 UT to 22:39:32 after the reset.
We produce the corresponding reconstructed sky images using a balanced
correlation [ref] with the mask pattern using FFT.  The detector image of
64 \x 64 pixels is rebinned into 8192 \x 8192 pixels in order to properly
scale the mismatch of the mask (4.7 mm) and detector pixel pitch
(5.1/2~mm)\footnote{The pixel pitch in a CZT crystal is 2.46 mm, but the gaps
between the detectors make the overall pixel pitch equivalent
to 2.55 mm (the detector pitch is 20.4 mm).}. 
Each sky image (every 12 s) is then rebinned to 768 \x 768
pixels covering down to the 10\% coding fraction, 
aspected corrected and then summed to produce the final image,
weighted by the coding fraction.

\subsection{Aspect Correction}

Fig.~\ref{timeline_cygx1}a shows the aspect and pointing 
history of the Cyg X-1 observation.
The star camera images are solved using astrometry.net \cite{Lang10}
after post-processing the raw images to remove several streaks due to a
defect in the CCD (see also \cite{Allen10}).  
The star camera images with a good aspect solution
were only available for a small fraction of the observation (5 images
before the reset around 21:15 UT and 63 images after the reset). For
the pre-reset data, most of which do not have a matching star camera image,
we correct the X-ray images using the azimuth and elevation given by
the DGPS calibrated magnetometer and the digital readout of the elevation axis with
shaft angle encoder (SAE) resolution of sub arcmin. For about 10
min of the usable post-reset data, we use the apect information by the
star camera images for the relative correction of the X-ray images.

\begin{figure}[t] \begin{center}
\footnotesize
\includegraphics*[width=0.47\textwidth]{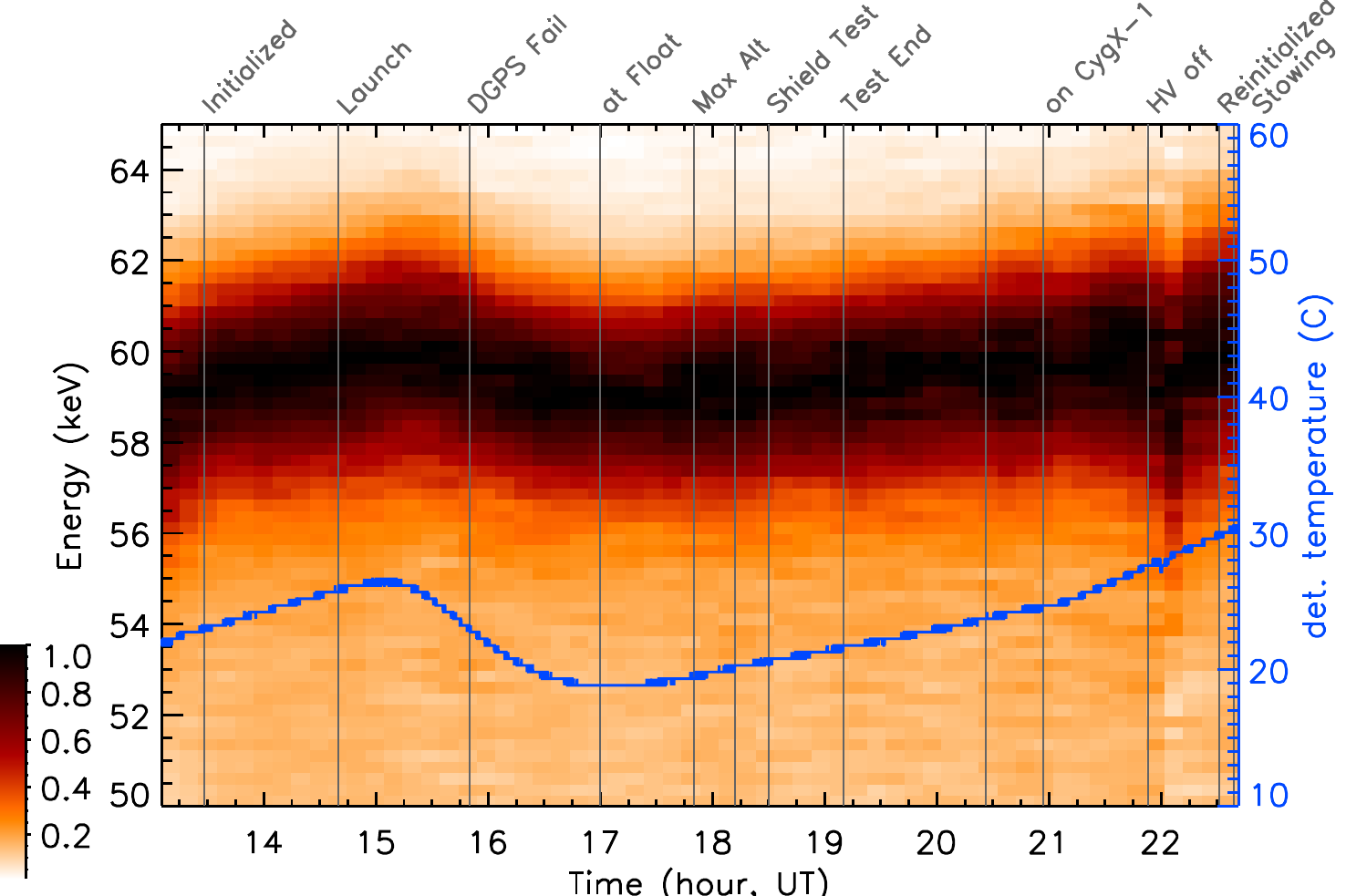}
\caption{Monitoring the gain variation
during the flight using AmCal sources. The normallized energy
histogram map as a function of time overlaid with the temperature of
the FCB. The gain change $\Delta E$ (keV) $\approx 0.116 \Delta T$ ({\Deg}C).
}
\label{amcal}
\end{center}
\end{figure}

Fig.~\ref{timeline_cygx1}b shows the relative offset between the
aspect of the star camera and the aspect information calculated by the
magnetometer and elevation axis. During the post-reset observation, the
relative offset is within 20\arcmin -- 30\arcmin in elevation and $\le 10$\arcmin
in azimuth, but there seems to be a long
term drift in the azimuth between the pre and post reset data, which can
be as large as 50\arcmin. This long term drift is the residual offset after
the DGPS calibration, and it originates from the azimuth hysteresis
of the magnetometer, which can be as large as a degree or two.
In order to compensate for this offset, the aspect of the X-ray image around
21:15 is corrected to match the post-reset data, and the rest of the
pre-reset data is corrected accordingly.

\begin{figure} \begin{center}
\footnotesize
\includegraphics*[width=0.47\textwidth]{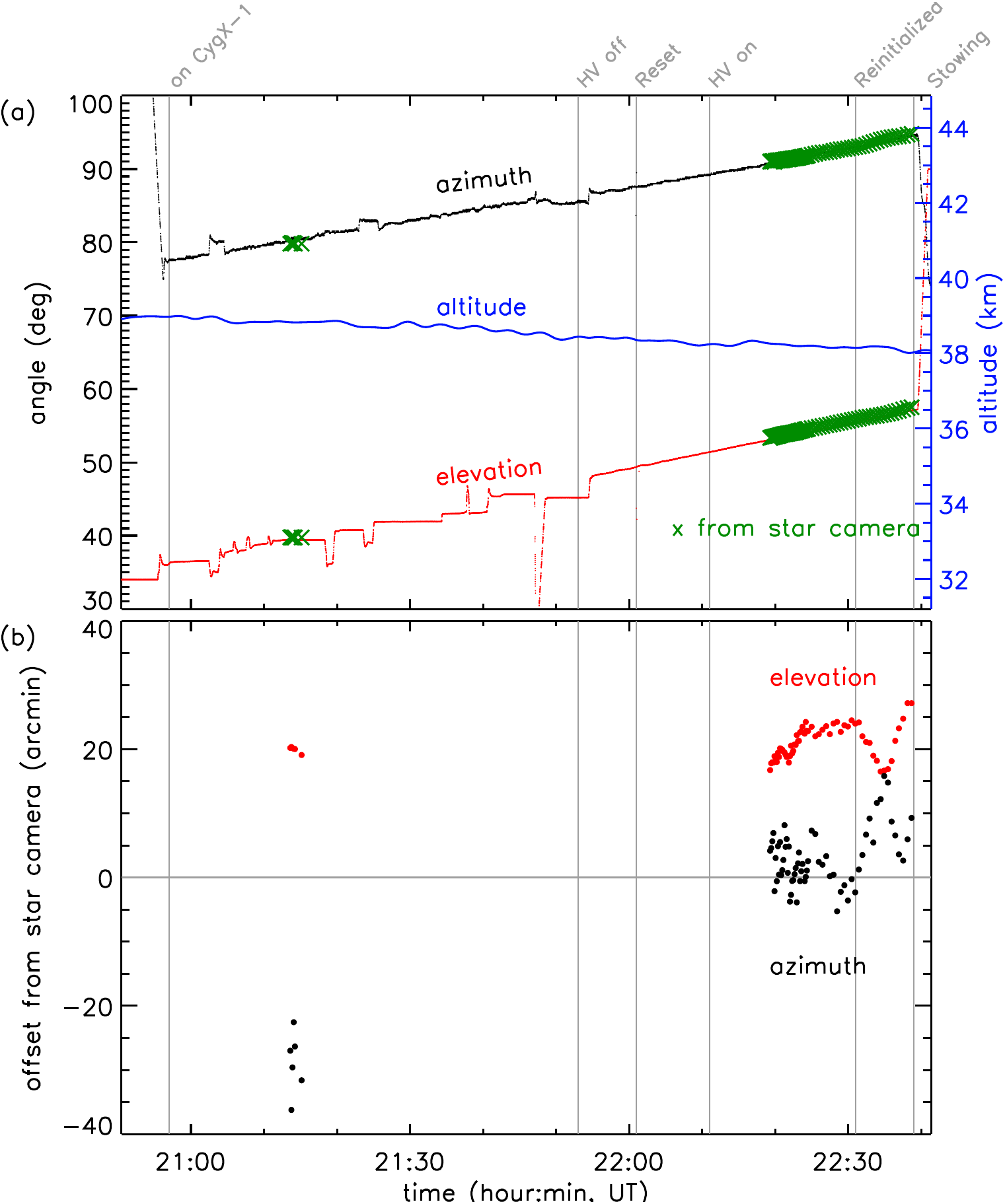}
\caption{(a) The aspect and pointing history and
(b) the relative offset of the aspect given by
the star camera vs.~aspect given by by the magnotometer and the
elevation axis shaft angle encoder during the observation of Cyg X-1.
The detector system was reset at around 22:00 UT for a test of in-flight
recylcing of power and initialization.
For imaging and spectral analysis of Cyg X-1,
the data taken from 20:57 to 21:53 UT (pre-reset: marked "on Cyg X-1" to
"HV off") and from  22:31 to 22:39 UT (post reset: marked "Reinitialized"
to "Stowing") were used.  }
\label{timeline_cygx1}
\end{center}
\end{figure}

A caveat of the aspect correction described above is that the offset
between the X-ray axis and the optical axis (or the axis of the star
camera) must be known. A delay in the payload integration left no time for
the X-ray vs.~star camera boresight measurement before the launch.  In order to resolve
an ambiguity in aspect correction due to the unknown boresight, we first
correct the pre-reset data according to the aspect information reported
by the magnetometer and the elevation axis only relative to the image at
21:15, assuming there is no X-ray boresight offset.  The combined image
shows a point source of about 6.5$\sigma$ at about 6\Deg off the center
of the X-ray image in the 30--100 keV band. Assuming this is CygX-1, one
can estimate the X-ray boresight, which was calculated to be 31\arcmin.
This X-ray boresight value is reasonable, given the uncertainties in
mounting the X-ray telescope and the star camera.

Assuming the 31\arcmin X-ray boresight offset, we re-apply the aspect
correction on the proper X-ray axis, and recombine the X-ray images.
In the recombined image of the pre-reset data, the same point source
now appeared at 6.8$\sigma$, indicating the proper boresighting has
increased the signal.  Finally we added the 10 min of the post-reset
data using the aspect information given by the star camera images.
In the final image, the point source appears at 7.2$\sigma$.
This solution gives Cyg X-1 at the expected location with a reasonable
boresight of the X-ray telescope to the
star camera. Fig.~\ref{image}a shows the reconstructed sky image in the 30
-- 100 keV band. The X-ray and optical axis are marked by '+' and 'x'.
The offset of Cyg X-1 from the X-ray axis is 6.4\Deg, resulting
in a partial coding fraction of 81\% and thus 19\% vignetting.

\begin{figure*} \begin{center}
\footnotesize
\includegraphics*[width=0.90\textwidth]{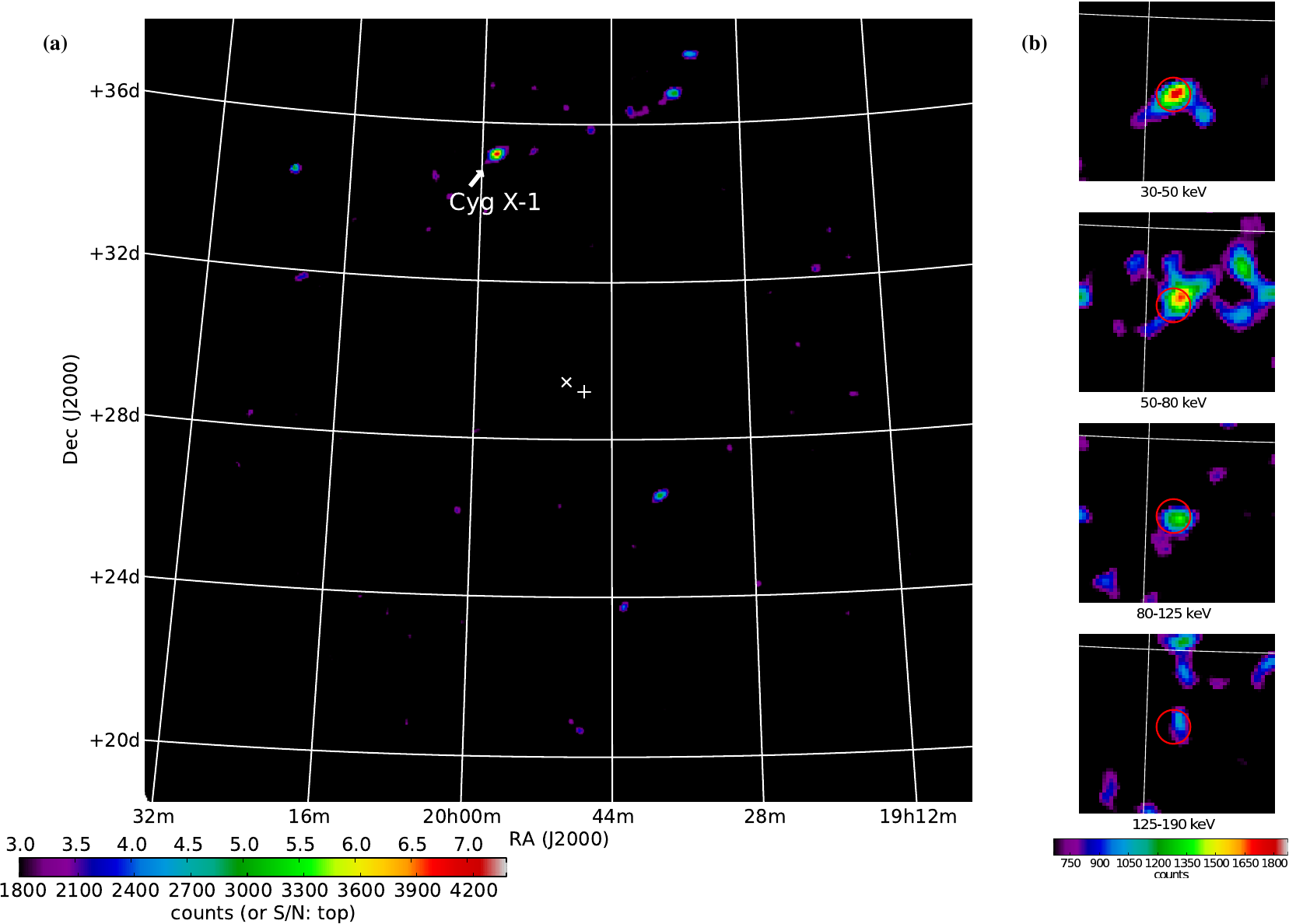}
\caption{Imaging Cyg X-1: (a) the reconstructed sky image in the 30--100
keV band above 60\% coding fraction.  (b) the 2\Deg \x 2\Deg thumbnail
images around Cyg X-1 in the 30--50, 50--80, 80--125 and 125--190
keV bands.  In (a), `+' indicates the center of the X-ray image, and
`x' indicates the star camera axis.  A $\sim$ 4$\sigma$ blob around
(R.A.,DEC.)=(19h 40m, 26\Deg) is a ghost image of Cyg X-1, caused by
the cyclic nature of the mask pattern (URA). The noise around (19h 36m,
37\Deg) is likely due to incomplete aspect correction, in particular,
large elevation jitters, since they are aligned along the elevation
direction from Cyg X-1.  The red circles in (b) show the angular
resolution (20\arcmin diameter).  }
\label{image}
\end{center}
\end{figure*}

\subsection{X-ray Spectrum of Cyg X-1}

In order to calculate the X-ray spectrum of Cyg X-1, we repeat the above
imaging procedure in the 30--50, 50--80, 80--125, and 125--190 keV bands.
The thumbnail images in Fig.~\ref{image}b show the reconstructed sky
images around Cyg X-1 in these energy bands.

Due to the relatively large uncertainties ($\sim$ 30\arcmin) of the aspect information
given by the magnetometer and the elevation axis,
a point source would appear somewhat broadened in the reconstructed sky
images.  For simplicity, we assume the source in the image is a point
source and so the maximum of the "blob" in the source image gives the detected source counts.
Using the maximum source counts is valid in our imaging procedure when the
background counts are much greater than the source counts.  In reality,
this will give us a lower bound for the source counts.

We calculate the source counts in each band, and apply a series of
corrections to get the incident rate, which include the average dead time
(45\%), the mask open fraction (38.7\%), the average coding fraction
(81\%), the average coded mask auto-collimation effect (2.6\%), the energy-dependent
QE and atmospheric absorption.  We assumed the QE of 86\% in the 60-100
keV range and for the rest of the energies, we use a linear interpolation
assuming 54\% at 30 keV (where only photo peaks are collected with a $\sim$
30 keV threshold) and 68\% at 200 keV (based on the total photon
cross-section of CZT).  For atmospheric absorption,
we use the atmospheric depth of 3.5 g cm\sS{-2} at the average altitude
38.5 km \cite{Stenov04} with the average elevation of 42.7\Deg.
In order to allow energy dependence, we use the attenuation of photons
in the atmosphere in \cite{Zombeck90} (p 203).
For both QE and atmospheric absorption, we calculate the energy bin averaged
estimates using a power law spectrum of photon index $\Gamma$=1.7.

The resulting spectrum is shown in
Fig.~\ref{spec}.  The error bars along the $x$ and $y$-axis represent the
energy bin size and the statistical errors respectively.
The red point shows the range of the source flux in 15--50 keV measured
by
\Swift/BAT\footnote{\url{http://heasarc.gsfc.nasa.gov/docs/swift/results/transients/}} 
within 12 hours of the \pe1 observation.  The closest BAT
observations are about 1--2 hours before or after the \pe1 observation.
The dashed line shows a representative hard
X-ray spectral model of Cyg X-1 in the literature \cite{Steinle82}. 
The observed flux is a bit lower than the other measurements but it is
certainly in an acceptable range,  given the high X-ray variability of
the source. The observed spectrum agrees with the known spectral
models within
3$\sigma$ and it can be fit by a simple power law model with photon
index $\Gamma$=1.7 $\pm$ 0.5 (the solid line). The reduced $\chi^2$
of the fit is small ($\sim 0.02$) due to the large uncertainties.
The observed spectrum is relatively harder, but 
the source detection
in the hard X-ray band above 125 keV is marginal
(Fig.~\ref{image}b).

\begin{figure} \begin{center}
\footnotesize
\raisebox{5ex-\height}{\includegraphics*[width=0.49\textwidth]{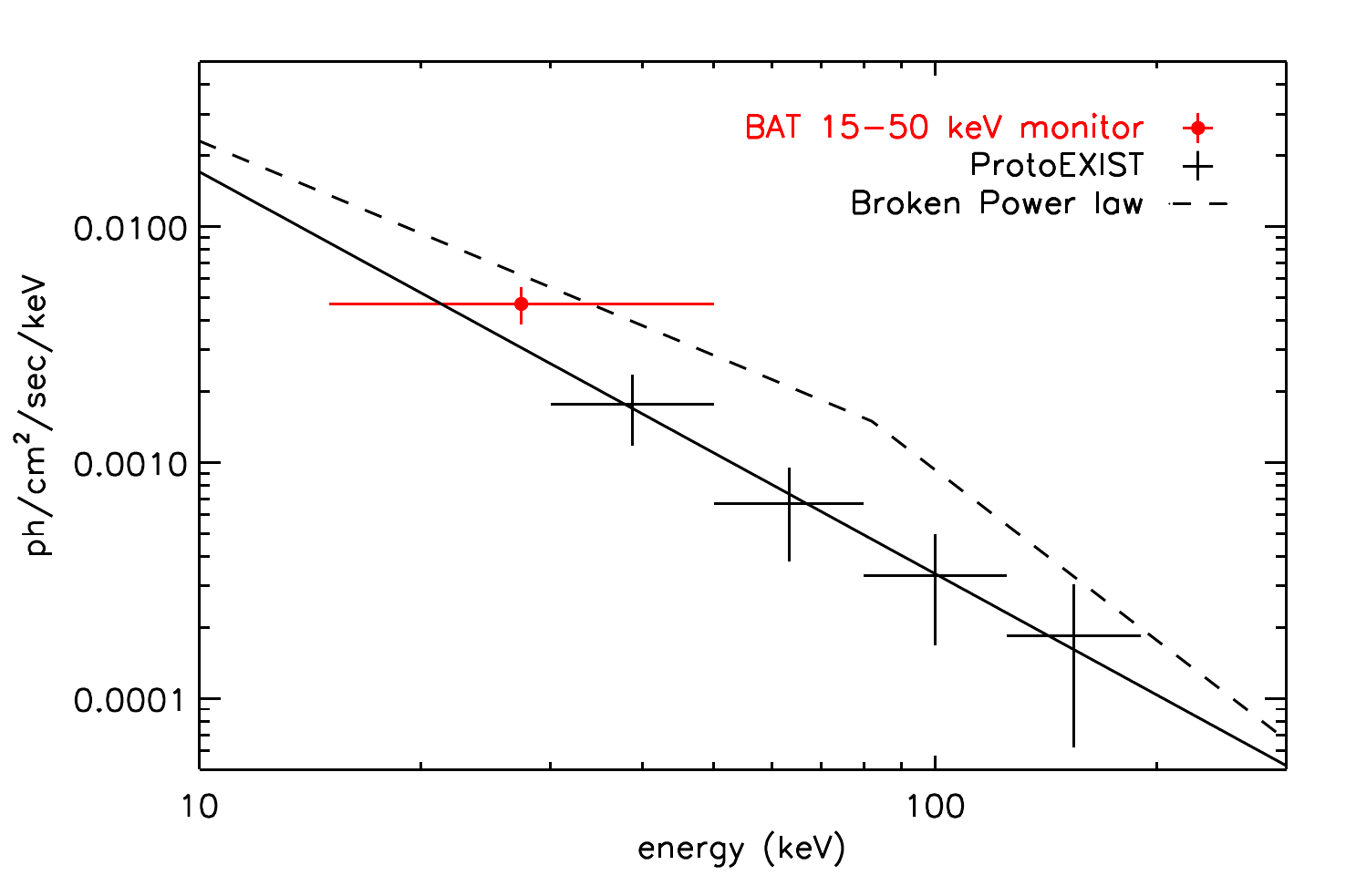}}
\caption{The observed hard X-ray spectrum of Cyg X-1. 
The (red) data point represents the range of the observed flux
in the 15--50 keV band by the BAT within 12 hours of the \pe1
observation.
The 
broken power model (dashed) is from Steinle, et al, 1982
\cite{Steinle82}. The solid line is a power law fit ($\sim E^{-\Gamma}$) 
to the data points, where the photon index $\Gamma$ = 1.7 $\pm$ 0.5.
}
\label{spec}
\end{center}
\end{figure}

\section{Summary and Future Development}

We have successfully carried out the first high altitude (40 km) balloon
flight of the wide-field hard X-ray telescope, \pe1, employing the first
generation of an advanced CZT imager. The CZT imager in \pe1 is,  to our knowledge,
the largest pixellated close-tiled CZT array at the moment (256
cm\sS{2} with 2.5 mm pixels).  The X-ray telescope performed excellently
throughout the 7.5 hr flight.  Despite a few problems in the new aspect
and pointing system, we were able to detect Cyg X-1 at 7.2$\sigma$ from
about an hour observation, $\sim$ 50 min of which
was with relatively poor aspect and so with smearing in the elevation
axis direction.  

Encouraged by this success, we have begun the next phase of the program,
\pe2, which will have 4\x finer detector spatial resolution 
(pixel pitch: 0.6 mm) and also both lower energy threshold
($\sim$ 5 -- 8 keV) and better energy resolution ($\sim$ 2 keV).  
To accomplish this,
a new ASIC, the Nu-ASIC, developed for
\NuSTAR\footnote{\url{http://www.nustar.caltech.edu/}} will be used in
the 8\x 8 DCU close-tiled CZT detectors of \pe2. The Nu-ASIC is developed
in the same ASIC family line as the RadNET ASIC, so they share a similar
basic operational architecture, which simplifies the transition. We will
combine the Nu-ASIC front-end readout electronics with a new packaging
and modularization architecture derived from the \pe1 system. For more
details of the \pe2 detector development, see also \cite{Hong10}.

\section{Acknowledgement} 

This work was supported by NASA grants NNG06WC12G and NNXO9AD76G to
Harvard University, and NASA grant NNX10AJ56G to Washington University
in St.~Louis.  We thank M.~Burke (SAO engineering), N.~Gehrels (GSFC),
K.~Dietz, C.~M.~Benson, B.~Ramsey (MSFC), D. Huie (Univ. of Alabama
Hunsville), W.~R.~Cook and F.~Harrison (Caltech) for their support in the
development and assembly of the \pe1 detector, telescope and gondola.
We also thank the McDonnell Center for the Space Sciences for their
support in the active CsI shield assembly, and the NASA CSBF balloon
launching team for their excellent support.

\bibliographystyle{elsart-num}

\end{document}